\documentclass[prd,preprint,aps,amsmath,amssymb,showpacs,superscriptaddress,nofootinbib]{revtex4-2}

\usepackage{graphicx} 
\usepackage{dcolumn}
\usepackage{bm}
\usepackage[colorlinks=true,citecolor=blue,urlcolor=blue,linkcolor=blue]{hyperref}
\usepackage{xcolor}
\usepackage[normalem]{ulem}
\usepackage{comment}
\usepackage{diagbox}
\usepackage{float}
\usepackage{url}

\newcommand{\MeV}{\rm MeV}
\newcommand{\GeV}{\rm GeV}

\begin{document}

\preprint{\hbox{PITT-PACC-2606}}

\title{Majoron Dark Matter, High-Scale Seesaw, and Leptogenesis}

\author{Brian Batell}
\email{batell@pitt.edu}
\affiliation{Pittsburgh Particle Physics, Astrophysics, and Cosmology Center, \\ Department of Physics and Astronomy, University of Pittsburgh, Pittsburgh, PA 15260, USA}
\author{Arnab Dasgupta}
\email{arnabdasgupta@pitt.edu}
\affiliation{Pittsburgh Particle Physics, Astrophysics, and Cosmology Center, \\ Department of Physics and Astronomy, University of Pittsburgh, Pittsburgh, PA 15260, USA}
\author{Swapnil Dutta}
\email{swd20@pitt.edu}
\affiliation{Pittsburgh Particle Physics, Astrophysics, and Cosmology Center, \\ Department of Physics and Astronomy, University of Pittsburgh, Pittsburgh, PA 15260, USA}
\author{Akshay Ghalsasi}
\email{aghalsasi@fas.harvard.edu}
\affiliation{Pittsburgh Particle Physics, Astrophysics, and Cosmology Center, \\ Department of Physics and Astronomy, University of Pittsburgh, Pittsburgh, PA 15260, USA}
\affiliation{Jefferson Physical Laboratory, Harvard University, Cambridge, MA 02138, USA}

\date{\today}

\begin{abstract}
We study the cosmology and observational probes of majoron dark matter in a high-scale seesaw framework with spontaneously broken lepton number. Right-handed neutrinos naturally generate light neutrino masses and can realize thermal leptogenesis, while the associated majoron is a light pseudo-Nambu-Goldstone boson that can be cosmologically stable and serve as a viable dark matter candidate for sub-MeV masses. We analyze both pre-inflationary and post-inflationary histories of lepton number breaking. In the pre-inflationary scenario, majoron dark matter is produced by misalignment and constrained by CMB isocurvature. 
In the post-inflationary scenario, the majoron abundance receives nonthermal contributions from spatially averaged misalignment, majoron radiation from global cosmic strings, and the collapse of the string-domain wall network, as well as a thermally produced component. 
This scenario can also be probed by future searches for the stochastic gravitational wave background produced by cosmic strings. 
We map the viable majoron dark matter parameter space and examine complementary probes from X-ray and soft $\gamma$-ray searches for majoron decays to photons, black hole superradiance, and Lyman-$\alpha$ forest observations. These results demonstrate that majoron dark matter offers a distinctive cosmological probe of high-scale lepton number breaking and thermal leptogenesis. 
\end{abstract}

\maketitle
\newpage

\section{Introduction}

The nonzero masses of neutrinos and the baryon asymmetry of the Universe provide two robust empirical indications of physics beyond the Standard Model (SM). A particularly elegant and economical framework addressing both is the high-scale type-I seesaw, in which the observed small active neutrino masses arise naturally from the exchange of heavy right-handed neutrinos (RHNs)~\cite{Minkowski:1977sc,Yanagida:1979as,Gell-Mann:1979ijt,Gell-Mann:1979vob,Glashow:1979nm,Mohapatra:1979ia,Schechter:1980gr}. 
If the heavy RHNs are thermally produced in the early Universe, their CP-violating, out-of-equilibrium decays can generate a lepton asymmetry that electroweak sphalerons partially convert into the observed baryon asymmetry, realizing thermal leptogenesis~\cite{Fukugita:1986hr}. This connection makes high-scale lepton number violation an attractive organizing principle, while motivating the question of how lepton number is broken and what cosmological and observational consequences follow.

A natural possibility is that lepton number is not simply explicitly broken at high scales, but is instead spontaneously broken~\cite{Chikashige:1980qk,Chikashige:1980ui,Gelmini:1980re}. In this case, the heavy Majorana RHN masses arise from the vacuum expectation value of a complex scalar field charged under lepton number, and the spectrum contains an associated pseudo-Nambu-Goldstone boson: the majoron. This light degree of freedom opens a new window onto the high-scale seesaw framework, which otherwise has few direct experimental consequences because the relevant dynamics occur at very high energies. 
In particular, the high scale of lepton number breaking can render the light majoron stable on cosmological timescales, opening the possibility that it addresses a third central empirical puzzle in particle physics and cosmology, namely, the particle nature of dark matter (DM). 
In this work, we study the cosmology and observational probes of majoron DM in a high-scale seesaw framework with spontaneously broken lepton number and thermal leptogenesis.

A key distinction in majoron cosmology is whether lepton number is broken during inflation and remains broken afterward, or is restored after inflation and breaks again during the subsequent thermal history.
This leads to two qualitatively distinct cosmological histories, analogous to the familiar pre-inflationary and post-inflationary scenarios of QCD axion DM (see Refs.~\cite{Marsh:2015xka,OHare:2024nmr} for a review). If lepton number is broken during inflation and not restored afterward, the majoron field takes a single coherent initial value throughout the observable Universe, and inflationary fluctuations of the majoron lead to isocurvature constraints. If lepton number is restored after inflation and spontaneously breaks later, the initial angle varies between causally disconnected patches, and the relic abundance receives contributions from misalignment as well as from the resulting string and domain wall network.

There are, however, important differences from the QCD axion case. The QCD axion potential is generated by nonperturbative QCD dynamics, which ties the axion mass to its decay constant. By contrast, the majoron mass depends on the source of explicit lepton number breaking and is generally independent of the majoron decay constant. In this work, we remain agnostic about the microscopic origin of this explicit breaking and study a minimal pseudo-Nambu-Goldstone boson potential with unit domain wall number. This provides a simple benchmark in the majoron mass--decay constant parameter space while avoiding the cosmological domain wall problem.

Majoron cosmology was discussed long ago in the context of Planck-suppressed breaking of global lepton number, with overclosure constraints already emphasized in Refs.~\cite{Akhmedov:1992hi,Cline:1993ht} and the majoron as a DM candidate considered in Refs.~\cite{Rothstein:1992rh,Berezinsky:1993fm}.
Subsequent studies have explored majoron DM in a variety of contexts, including connections to inflation~\cite{Kazanas:2004kv,Boucenna:2014uma}, decaying warm DM~\cite{Lattanzi:2007ux}, indirect detection probes~\cite{Bazzocchi:2008fh,Lattanzi:2013uza,Garcia-Cely:2017oco,McKeen:2018xyz}, neutrino flavor models~\cite{Esteves:2010sh}, freeze-in production~\cite{Heeck:2017xbu,Manna:2022gwn,King:2024idj}, and production from topological defects~\cite{Reig:2019sok,Greljo:2025suh}. For further studies on related topics, see Refs.~\cite{Gu:2010ys,Frigerio:2011in,Queiroz:2014yna,Ibe:2015nfa,Rojas:2017sih,Shakya:2018qzg,Brune:2018sab,Heeck:2019guh,Biggio:2023gtm,deGiorgi:2023tvn,Akita:2023qiz,Chun:2025abp,Chao:2022blc,Chao:2023ojl,Peng:2025sri,Long:2017rdo,Berbig:2025hlc,Falkowski:2017uya,Ghosh:2025cxp,Chianese:2025mll}.

We begin in Section~\ref{sec:model} by introducing the singlet majoron model, reviewing the necessary conditions  for high-scale leptogenesis, and discussing majoron decays, showing that sub-MeV majorons can be cosmologically long-lived. Next, in Section~\ref{sec:pre-inf} we then study majoron DM in the pre-inflationary scenario, focusing on misalignment production and CMB isocurvature constraints. Then we turn in Section~\ref{sec:post-inf} to the post-inflationary scenario, where misalignment, majoron radiation from cosmic strings, string-domain wall network collapse, and possible thermal production all contribute to the relic abundance. 
We also estimate the stochastic gravitational wave spectrum from cosmic strings in this scenario. 
For both cosmological histories, we also examine complementary probes from X-ray and soft $\gamma$-ray searches for majoron decays to photons, black hole superradiance, and Lyman-$\alpha$ forest observations. Finally, we summarize our results and conclude in Section~\ref{sec:Conclusions}. 
We also provide some technical details on the treatment of anharmonic effects in Appendix \ref{app:Fcorrection}.

\section{Singlet Majoron Model}
\label{sec:model}

In this section we begin by introducing the singlet majoron model~\cite{Chikashige:1980qk,Chikashige:1980ui,Gelmini:1980re}. 
The Lagrangian of the lepton sector contains the following interactions:
\begin{align}
    \label{eq:Lagrangian}
    -\mathcal{L} &  \supset  \overline L y N_R H + \frac{1}{2} \sigma \overline N_{\! R}^{\,c} \lambda N_R  + {\rm h.c.} \, .
\end{align}
Here $N_R$ denotes the 
three generations of RHNs with lepton number $+1$, $L$ are the $SU(2)_L$ doublet leptons which carry lepton number $+1$, $\sigma$ is a gauge singlet complex scalar with lepton number $-2$, $H$ is the Higgs doublet field, and $y$ and $\lambda$ denote the Yukawa couplings involving $H$ and $\sigma$, respectively. 
By applying suitable flavor transformations we are free to work in a basis in which $\lambda$ is diagonal and real. We also assume the existence of a suitable scalar potential 
which causes $H$ and $\sigma$ to obtain VEVs at zero temperature, $\langle H \rangle = (v/\sqrt{2}, 0)^T$ and $\langle \sigma \rangle = f_\phi/\sqrt{2}$ with $v = 246$ GeV and $v \ll f_\phi$.

At high temperatures, $v \ll T \lesssim f_\phi$, relevant for leptogenesis and cosmological majoron production, lepton number is spontaneously broken due to the VEV of $\sigma$, while  electroweak symmetry is unbroken. 
In this phase, the RHNs obtain masses $M_R = \lambda f_\phi/\sqrt{2}$.  
The complex scalar field $\sigma$ is parameterized as
\begin{equation}
    \sigma = \frac{f_{\phi}}{\sqrt{2}} e^{i \phi/f_{\phi}}, 
\end{equation}
where $\phi$ is the majoron, the pseudo-Nambu-Goldstone boson associated with spontaneous lepton number breaking.\,\footnote{We assume the radial mode of $\sigma$ is heavy and can be neglected for the purpose of determining the final majoron abundance. Although radial mode decays can produce majorons in the post-inflationary scenario, this population thermalizes, so its abundance is set by standard thermal equilibrium considerations. See Section~\ref{sec:therm} for further details.}
We further assume the existence of terms that explicitly break lepton number and generate a potential for the majoron field. While we are agnostic about their origin, such terms could arise, for instance, from quantum gravity effects~\cite{Akhmedov:1992hi,Rothstein:1992rh,Alonso:2017avz}.
Under these assumptions, the majoron effective Lagrangian below the lepton number breaking scale $f_\phi$ contains the following kinetic and potential terms 
\begin{align}
    \label{eq:majoron}
    \mathcal{L}  \supset \frac{1}{2}\partial_{\mu} \phi \partial^{\mu}\phi
    - m^{2}_{\phi} f_{\phi}^{2}\left[1- \cos(\phi/f_\phi)\right] \, ,
\end{align}
where $m_{\phi}$ is the mass the majoron acquires through explicit breaking of $U(1)_L$. For later convenience, we introduce the angular variable $\theta = \phi/f_{\phi}$. 
We will use this effective Lagrangian, Eq.~(\ref{eq:majoron}), to study the cosmological properties of the majoron, including its relic abundance and isocurvature perturbations. 

The potential in Eq.~\eqref{eq:majoron} corresponds to the minimal case with domain wall number $N_{\rm DW}=1$, meaning that there is a single minimum within the original $2\pi$ field range of the angular variable $\theta=\phi/f_\phi$. The analogous scenario has been studied extensively for axions~\cite{Sikivie:1982qv,Vilenkin:1982ks,Chang:1998tb,Hiramatsu:2012gg}, and we expect the qualitative conclusions to apply also to majorons. More general explicit symmetry-breaking potentials can contain $N_{\rm DW}>1$ degenerate minima associated with a discrete $Z_{N_{\rm DW}}$ symmetry. In this work we focus on the minimal $N_{\rm DW}=1$ case. The qualitative implications of $N_{\rm DW}>1$ for cosmology, including the possible appearance of stable domain wall networks, their dependence on additional bias terms, and the changes to majoron and gravitational wave production, are briefly discussed in Secs.~\ref{sec:strings} and \ref{GWIsoPostInf}.

At even lower temperatures, $T \lesssim v$, we must account for the effects of electroweak symmetry breaking, which are relevant for the decays of the majoron, and for this we follow closely the analyses of Refs.~\cite{Garcia-Cely:2017oco,Heeck:2019guh}. 
Following electroweak symmetry breaking, mass terms are generated for the SM neutrinos,
\begin{align}
\label{eq:neutrino-mass}
 -{\cal L} & \supset \frac{1}{2} \biggl(\overline \nu_L ~~ \overline N_{\! R}^{\,c} \biggr)
 \left( 
  \begin{array}{cc}
  0 & M_D \\
  M_D^T & M_R
  \end{array}
 \right)
 \left(  
 \begin{array}{c}
 \nu_L^c  \\
 N_R 
 \end{array}
 \right) +{\rm h.c.}  \\
& = \frac{1}{2} \, \overline n_{\! R}^{\,c} \, V^T
\left(
 \begin{array}{cc}
 0 & M_D \\
 M_D^T & M_R
 \end{array}
\right)
 V n_R+{\rm h.c.} 
\nonumber \\
 & = \frac{1}{2} \, \overline n_{\! R}^{\,c} \, M_n \, n_R + {\rm h.c.} \, ,
 \end{align}
 where $M_D = y v/\sqrt{2}$ and $M_R = \lambda f_{\phi}/\sqrt{2}$ (defined already above) are the Dirac mass matrix and RHN Majorana mass matrix, respectively, while 
$M_n = \operatorname{diag}(m_{\nu_1},\allowbreak
m_{\nu_2},\allowbreak
m_{\nu_3},\allowbreak
m_{N_1},\allowbreak
m_{N_2},\allowbreak
m_{N_3})$
 is the diagonalized neutrino mass matrix. The change from the flavor basis to the mass basis is accomplished by a unitary transformation, $(\nu_L^c ~ N_R )^T =  V n_R$.
The $6 \times 6$ mixing matrix $V$ can be written as 
\begin{align}
\label{eq:V}
V 
& \simeq
\left(
 \begin{array}{cc}
 U^{*} & -i U^{*} \sqrt{d_{l}} R^{\dag} \sqrt{d^{-1}_{h}} \\
 -i \sqrt{d^{-1}_{h}} R \sqrt{d_{l}} & 1
 \end{array}
\right),
\end{align}
where $U$ is the Pontecorvo–Maki–Nakagawa–Sakata (PMNS) matrix~\cite{Maki:1962mu,Pontecorvo:1967fh}, 
$d_{l}$ ($d_{h}$) is the diagonal mass matrix of the light, mostly active (heavy, mostly sterile) neutrinos, and $R$ is a complex orthogonal matrix. 
The neutrino Yukawa coupling $y$ can then be written as 
\begin{align}
    \label{eq:CIS}
    y = \frac{i \sqrt{2}}{v}  U \sqrt{d_{l}} R^{T} \sqrt{d_{h}} \, ,
\end{align}
which is the familiar Casas-Ibarra parametrization~\cite{Casas:2001sr}. 

The majoron couples to RHNs through the  $\tfrac{1}{2} \sigma \overline N_{\! R}^{\,c} \lambda N_R$ term in Eq.~\eqref{eq:Lagrangian}. Neutrino mixing then induces tree-level couplings to neutrino mass eigenstates and effective one-loop couplings to SM fermions through weak interactions. At two loops, the majoron also couples to pairs of gauge bosons. 
In Section~\ref{sec:decays} below, we summarize the decay channels and rates relevant for this work, following the analyses in Refs.~\cite{Garcia-Cely:2017oco,Heeck:2019guh}.

Since quantum gravity is expected to violate global symmetries, the majoron mass can receive contributions from Planck-suppressed operators that break lepton number, $\Phi^n / {M_{\rm pl}^{n-4}}+{\rm h.c.}$, parametrically given by  
\begin{align}
    \label{eq:majoron_mass}
    \delta m_{\phi} & \sim n M_{\rm pl} \left(\frac{f_{\phi}}{M_{\rm pl}}\right)^{\frac{n-2}{2}}  \, .
\end{align} 
For light majorons and large $f_{\phi}$, naturalness requires the leading lepton number violating operator to have large enough dimension $n$ so that the induced correction satisfies $\delta m_\phi \lesssim m_\phi$.
This potentially poses a quality problem analogous to the QCD axion quality problem. Various approaches to address the axion quality problem have been proposed in literature (see Ref.~\cite{DiLuzio:2020wdo} for a review) and can conceivably be adapted to address the majoron quality problem. Moreover, high-quality majorons may arise naturally in extra-dimensional or string-theoretic constructions, through mechanisms analogous to those that protect the quality of QCD axions~\cite{Svrcek:2006yi,Choi:2003wr,Reece:2025thc,Catinari:2024zon,Petrossian-Byrne:2025mto}. In this work, we remain agnostic about the microscopic origin of the explicit breaking and treat the majoron mass as a free parameter in the potential of Eq.~\eqref{eq:majoron}.

\subsection{Leptogenesis}
\label{sec:lepto}
We are interested in regions of parameter space where the singlet majoron model can be compatible with the observed baryon-to-photon ratio through thermal leptogenesis. In this scenario, CP-and lepton number-violating decays of the thermally produced heavy RHNs generate a lepton asymmetry, which is subsequently converted into a baryon asymmetry by electroweak sphalerons. 
Here we briefly review the standard vanilla leptogenesis estimate of the baryon asymmetry and the resulting implications for our scenario.

The baryon-to-photon ratio in vanilla thermal leptogenesis can be parameterized as~\cite{Buchmuller:2004nz}
\begin{align}
    \label{eq:YB}
    \frac{n_{B}}{n_{\gamma}} & \simeq 0.96 \times 10^{-2} \, |\epsilon_{1}| \, \kappa_{f}(K) \, ,
\end{align}
 where $K$ is the decay parameter defined below in Eq.~(\ref{eq:tildem1}).
This should be compared to the observed value, $n_{B}/n_{\gamma}\vert_{\rm obs} = 6 \times 10^{-10}$~\cite{Planck:2018vyg,Yeh:2026pil}. The leptogenesis prediction for the baryon asymmetry, Eq.~(\ref{eq:YB}), depends on two key quantities: 1) $\epsilon_1$, the CP-violating decay asymmetry parameter of the lightest RHN, quantifying the difference between the rates for decays into leptons and antileptons, and 2) $\kappa_{f}(K)$, an efficiency factor that encodes the net effect of $N_1$ production and washout processes on the final lepton asymmetry.\footnote{The numerical prefactor $(0.96 \times 10^{-2})$ accounts for both the sphaleron conversion of the $B-L$ asymmetry into a baryon asymmetry and entropy dilution between leptogenesis and recombination.}

The final lepton asymmetry depends on both RHN decays/inverse decays and scattering processes in the thermal bath, which in turn are governed by the Yukawa couplings of $N_1$. The rates for these processes are conveniently parameterized in terms of  the effective neutrino mass parameter, $\tilde m_1$, and decay parameter, $K$, which are respectively given by~\cite{Buchmuller:2004nz}
\begin{align}
    \label{eq:tildem1}
    \tilde{m}_{1} &\equiv \frac{v^{2}~\left(y^{\dag} y\right)_{11}}{2 m_{N_{1}}} = \left(m_{\nu_1} |R_{11}|^{2} + m_{\nu_2} |R_{12}|^{2} + m_{\nu_3} |R_{13}|^{2}\right)\nonumber ,\\
    K &\equiv \frac{\left(y^{\dag} y\right)_{11} M_{\rm pl}}{ \gamma m_{N_{1}}} = \frac{2 \tilde{m}_{1} M_{\rm pl}}{\gamma v^{2}} = \frac{\tilde m_{1}}{ 1.1 \times 10^{-3} \,
    {\rm eV}}, 
\end{align}
where $M_{\rm pl} = 1.22 \times 10^{19} \, \GeV$ is the Planck mass  and 
$\gamma = 16 \pi^{5/2} g^{1/2}_{*}/(3 \sqrt{5}) \simeq 431$ 
for $g_{*} = 106.75$. We have used the Casas-Ibarra parametrization (\ref{eq:CIS}) to express $\tilde{m}_{1}$ in terms of the active neutrino masses. 
The decay parameter $K$ is the ratio of the $N_1$ decay rate to Hubble parameter at $T = m_{N_{1}}$ and thus characterizes the strength of the $N_1$ interactions with the plasma. 
If $K \ll 1$, $N_1$ is not efficiently thermally populated, suppressing the final lepton asymmetry. 
If $K \gg 1$, inverse decays and related washout processes are efficient, which can erase the lepton asymmetry generated by $N_1$ decays.
From Eq.~(\ref{eq:tildem1}), assuming  $m_{\nu_1} \ll m_{\nu_{2,3}}$ and generic $R_{ij} \sim {\cal O}(1)$, we expect $\tilde{m}_{1} \sim \sqrt{|\Delta m^2_{\rm atm}|} \simeq 0.05$ eV which gives $K\sim \mathcal{O} (10)$. 
The eventual baryon asymmetry is therefore a function of $K$, encoded in the efficiency factor $\kappa_{f}(K)$ in Eq.~(\ref{eq:YB}). 

Consider next the asymmetry parameter $\epsilon_1$ in Eq.~(\ref{eq:YB}). 
Assuming a hierarchical RHN spectrum, $m_{N_{1}} \ll m_{N_{2,3}}$, this can be written as~\cite{Covi:1996wh,Davidson:2002qv}
\begin{align}
    \label{eq:epsilon}
    \epsilon_{1} & \simeq -\frac{3}{16 \pi (y^{\dagger}y)_{11}} \sum_{j = 2,3} {\rm Im}\left[\left(y^{\dagger} y\right)^2_{1j}\right] \frac{m_{N_{1}}}{m_{N_{j}}} 
    = \frac{3}{8\pi} \frac{m_{N_1}}{v^2} \frac{\sum_i m^2_{\nu_i} {\rm Im}(R_{1i}^2)}{\sum_i m_{\nu_i} |R_{1i}|^2 }.
\end{align}
Using the orthogonality of $R$, one finds that the maximum CP asymmetry is controlled by the difference between the largest and smallest light neutrino masses. Since this difference cannot exceed the atmospheric mass scale, $|\Delta m^2_{\rm atm}| = 2.5 \times 10^{-3}\,{\rm eV}^2$~\cite{Esteban:2024eli}, the asymmetry is bounded from above~\cite{Davidson:2002qv}, 
\begin{align}
    \label{eq:epsilon-bound}
    |\epsilon_{1}| & \lesssim  \frac{3}{8 \pi} \frac{m_{N_1} \sqrt{|\Delta m^2_{\rm atm}|}}{v^{2}} \simeq  10^{-7} \left(\frac{m_{N_{1}}}{10^{9} \,{\rm GeV}}\right).
\end{align}

The efficiency factor takes a maximum value of $\kappa_{f, \rm max} \simeq 0.2$ at an intermediate value $K \sim {\cal O}(1)$. 
Therefore, to explain the observed baryon-to-photon ratio, we require $|\epsilon_{1}| \gtrsim 3 \times 10^{-7}$. Accounting for the maximum size of $|\epsilon_1|$ in Eq.~(\ref{eq:epsilon-bound}), one obtains a lower bound on the mass of the lightest RHN, $m_{N_1} \gtrsim 3 \times 10^9$ GeV. This is the well-known Davidson-Ibarra bound on the lightest RHN mass in vanilla thermal leptogenesis~\cite{Davidson:2002qv}.   

The Davidson-Ibarra bound has important implications for our study. First, since we are considering spontaneous lepton number breaking,  $m_{N_i} = \lambda_i f_\phi/\sqrt{2}$, and restricting to perturbative couplings $\lambda_i$, we impose the rough boundary $f_\phi \gtrsim 10^{9}$ GeV. 
This implies that the majoron is feebly coupled to the RHNs and SM particles and can therefore be stable on cosmological time scales. Second, successful thermal leptogenesis requires the reheating temperature to be similar to or larger than the lightest RHN mass, $T_{\rm RH} \gtrsim m_{N_1}$, which we will impose throughout this work. 
As is familiar from the case of the QCD axion, the cosmological properties of the majoron further depend on whether the lepton number symmetry was broken during inflation/reheating or not, referred to as the {\it pre-inflationary} and {\it post-inflationary} scenarios, respectively. We will consider $m_{N_1} \lesssim T_{\rm RH}  \lesssim  f_\phi$ 
for the pre-inflationary scenario and $m_{N_1} \lesssim f_\phi \lesssim  T_{\rm RH}$ for the post-inflationary scenario. 

We emphasize that the discussion above neglects several potentially important ingredients, including flavor effects~\cite{Nardi:2006fx,Abada:2006ea}, contributions from the heavier RHNs, thermal corrections~\cite{Giudice:2003jh}, and resonant enhancement from quasi-degenerate RHNs~\cite{Pilaftsis:2003gt}. As is well known, these can modify the leptogenesis prediction and allow for lighter RHNs, but their inclusion lies beyond the scope of this work. The large majoron decay constants implied by vanilla thermal leptogenesis imply that the majoron has extremely feeble couplings to the SM particles, which will allow it to be cosmologically long lived, as we discuss next. 

\subsection{Majoron Decay}
\label{sec:decays}

Since our focus is on the majoron as a DM candidate, a basic requirement is that it be stable on cosmological time scales, and in particular sufficiently long-lived to satisfy astrophysical bounds on decaying DM. 
We must therefore identify its dominant decay channels, which depend on the majoron mass, decay constant, and the neutrino Yukawa coupling.

As discussed in the previous subsection, viable thermal leptogenesis points to decay constants $f_\phi \gtrsim 10^{9}$ GeV, which allows for $N_1$ to satisfy the Davidson-Ibarra bound for perturbative couplings $\lambda$. 
We also restrict to $f_\phi \leq 10^{16}\,{\rm GeV}$, maintaining a modest hierarchy between the lepton number breaking scale and the Planck scale.\footnote{A rough upper bound of $f_\phi \lesssim 10^{16}$ GeV can also be motivated by bounds on the inflationary Hubble scale from the non-observation of tensor modes in the CMB; see Section~\ref{sec:pre-inf-other}.} 
Along with the  decay constant $f_\phi$, the majoron couplings depend on the neutrino Yukawa couplings, which are determined by the PMNS matrix elements, the light neutrino and heavy RHN masses, and the $R$ matrix via Eq.~(\ref{eq:CIS}). 
The PMNS matrix and the neutrino mass-squared splittings have been determined through global fits to a variety of neutrino oscillation measurements~\cite{Esteban:2024eli,ParticleDataGroup:2024cfk}. 
We furthermore impose consistency with cosmological constraints on the sum of neutrino masses~\cite{Naredo-Tuero:2024sgf}.
Current neutrino data allow the lightest active neutrino to be massless, and in this limit, the lightest RHN can saturate the Davidson-Ibarra bound in Eq.~\eqref{eq:epsilon-bound}. 
Additionally, we note that our main results for majoron cosmology are not substantially affected by the choice of neutrino mass ordering.
Finally, we will consider two cases for the typical size of the $R$ matrix elements: 1) $|R_{ij}| \sim {\cal O}(1)$, and 2) larger values that satisfy the perturbativity of the Yukawa couplings, ${\rm max}(y_{ij}) < \sqrt{4\pi}$. 

Majorons heavier than the MeV-scale and below the muon mass decay dominantly to electron-positron pairs. The partial decay rate is given by~\cite{Garcia-Cely:2017oco} 
\begin{align}
    \label{eq:majoron_decay}
    \Gamma_{\phi \rightarrow ee} = \frac{|g_{\phi ee}|^2 m_{\phi}}{8\pi} \left(1-\frac{4m_e^2}{m_\phi^2}\right)^{1/2} \, ,
\end{align}
where $g_{\phi ee}$ is the one-loop majoron-electron coupling,
\begin{align}
    \label{eq:gPaee}
    g_{\phi ee} = \frac{m_{e}}{8 \pi^{2} v} \left(-\frac{1}{2}{\rm Tr}\,\mathcal{K} + \mathcal{K}_{ee}\right) \, ,
\end{align}
with
\begin{align}
    \label{eq:K}
    \mathcal{K} = \frac{M_{D} M^{\dagger}_{D}}{v f_{\phi}} = \frac{1}{v f_{\phi}} U \sqrt{d_{l}} R^{T} d_{h} R^{*} \sqrt{d_{l}} U^{\dagger} \, .
\end{align}
The second step in the last equation follows from the use of Eq.~(\ref{eq:CIS}). 

To estimate the longest typical majoron decay lifetime to electrons and corresponding majoron lifetime, we take $R_{ij} \sim {\cal O}(1)$, in which case $|\mathcal{K}_{ij}| \sim \sqrt{\Delta m^2_{\rm atm}} m_{N_3}/(v f_\phi)$ and we obtain,
\begin{equation}
\label{eq:lifetime-phi-ee}
\tau_{\phi\rightarrow ee} \lesssim  10^{28} \, {\rm s} \times  \left(\frac{2\, {\rm MeV}}{m_\phi}\right) 
\left(\frac{10^{11}\, {\rm GeV}}{m_{N_3}}\right)^2
\left(\frac{f_\phi}{10^{15}\, {\rm GeV}}\right)^2.
\end{equation}
Although the majoron lifetime in this regime is much longer than the age of the Universe, $\tau_{U} = 4\times 10^{17} \,{\rm s}$, 
much stronger limits apply if majorons make up all of the DM. In particular, searches for $\gamma$-ray signals from electron-positron annihilation constrain DM decays to $e^+ e^-$ to have lifetimes
$\tau_{\phi\rightarrow ee} \gtrsim 10^{29} \, {\rm s}$~\cite{DelaTorreLuque:2023cef}.
The estimate in Eq.~\eqref{eq:lifetime-phi-ee} implies that it is challenging for a majoron with mass $m_\phi \geq 2m_e$ to be all of the DM while remaining compatible with vanilla thermal leptogenesis, unless the majoron mass lies relatively close to twice the electron massm, or $f_\phi \sim \mathcal{O}(10^{16}\,{\rm GeV})$, or there are accidental cancellations or additional symmetries that further suppress $g_{\phi ee}$.
Although we do not explore this possibility further here, a dedicated study of whether future $\gamma$-ray observatories could rule out majoron DM in this small region of parameter space would be worthwhile.

For sub-MeV majorons, the dominant decay channels are into neutrino pairs and photon pairs.
The partial decay rate to neutrinos 
is given by \cite{Garcia-Cely:2017oco} 
\begin{equation}
\Gamma_{\phi \rightarrow \nu \nu}  = \frac{m_\phi}{16\pi f_\phi^2} \sum_i m_{\nu_i}^2 \, .
\end{equation}
Since $\sum_i m_{\nu_i}^2 \approx \Delta m_{\rm atm}^2$ assuming a normal hierarchy and massless lightest neutrino, we obtain 
\begin{equation}
    \label{eq:tauanunu}
    \tau_{\phi\rightarrow\nu\nu} \simeq  10^{33}\,\textrm{s}\,\left(\frac{\textrm{MeV}}{m_\phi}\right){\left(\frac{f_\phi}{10^{16}\textrm{ GeV}}\right)}^2 \, .
\end{equation}
We see that sub-MeV majorons are cosmologically long-lived. The neutrino decay channel $\phi\to\nu\nu$ is difficult to probe indirectly in this mass range, and existing neutrino telescope searches are not expected to provide competitive constraints; see, e.g., Refs.~\cite{Akita:2023qiz,Arguelles:2022nbl,McKeen:2018xyz}.

The majoron can also decay to a pair of photons at two loops \cite{Heeck:2019guh}. The decay rate to photons is given by 
\begin{equation}
\Gamma_{\phi \rightarrow \gamma \gamma} = \frac{|g_{\phi\gamma\gamma}|^2 m_\phi^3}{64 \pi} \, ,
\end{equation}
where $g_{\phi\gamma\gamma}$ is the effective majoron-photon coupling. For $m_\phi \ll 2 m_e$, the electron contribution dominates and  the majoron-photon coupling can be expressed in terms of the majoron-electron coupling, 
$|g_{\phi\gamma\gamma}| \simeq |g_{\phi ee}| \, \alpha \, m_\phi^2 \, /(12\,\pi \, m_e^3)$,  with $g_{\phi ee}$ given in Eq.~(\ref{eq:gPaee}). Being a two-loop process, the majoron decay to photons is naturally suppressed, but could still potentially be observable for large neutrino Yukawa couplings. To estimate the maximum possible decay rate, we saturate the perturbativity bound ${\rm max}(y_{ij}) < \sqrt{4\pi}$, implying $\mathcal{K}_{ij} \lesssim 2\pi v/f$. This gives an estimate for the lifetime
\begin{equation}
\label{eq:tauagg}
\tau_{\phi\rightarrow\gamma\gamma}\gtrsim 10^{26}\textrm{ s} ~ {\left(\frac{0.1\textrm{ MeV}}{m_\phi}\right)}^7 {\left(\frac{f_\phi}{10^{12}\textrm{ GeV}}\right)}^2  \hspace{5pt}.
\end{equation}

For $m_\phi\lesssim \MeV$, majoron decays to photons can be probed by soft $\gamma$-ray and X-ray searches. We present the corresponding constraints from Eq.~\eqref{eq:tauagg} in Secs.~\ref{sec:pre-inf} and~\ref{sec:post-inf}.

Thus, in the sub-MeV mass range, the majoron is cosmologically long-lived and can serve as a viable DM candidate. In the next two subsections, we study its cosmological production and associated probes in the pre-inflationary and post-inflationary scenarios.

\section{Majoron dark matter in the pre-inflationary scenario}
\label{sec:pre-inf}

We first consider the {\it pre-inflationary} scenario in which lepton number is spontaneously broken during inflation, $H_I/(2\pi) < f_\phi$ with $H_I$ is the inflationary Hubble parameter, and remains broken afterward because the plasma never restores the symmetry. 
This is typically ensured by a sufficiently low reheating temperature, $T_{\rm RH}\ll f_\phi$, although more precisely the maximum temperature attained after inflation must remain below the $U(1)_L$ symmetry breaking scale.
In this scenario, the observable Universe originates from a single inflated patch, so the majoron angle $\theta_i = \phi_i/f_\phi$ is approximately homogeneous across our present horizon, up to inflationary fluctuations. If the majoron is sufficiently light, $m_\phi \lesssim \mathcal{O}({\rm MeV})$, it is cosmologically long-lived (see Section~\ref{sec:decays}) and can account for the observed DM abundance via the misalignment mechanism. 
Realizing this scenario together with standard thermal leptogenesis imposes additional requirements on the hierarchy of scales. The reheating temperature must be high enough to thermally populate the lightest RHN but low enough that the lepton number symmetry is not thermally restored after inflation, leading parametrically to the conditions $m_{N_1}\lesssim T_{\rm RH}\lesssim f_\phi$.

\subsection{Majoron abundance from misalignment production}

\begin{figure}
    \centering
    \includegraphics[width = 0.9\linewidth]{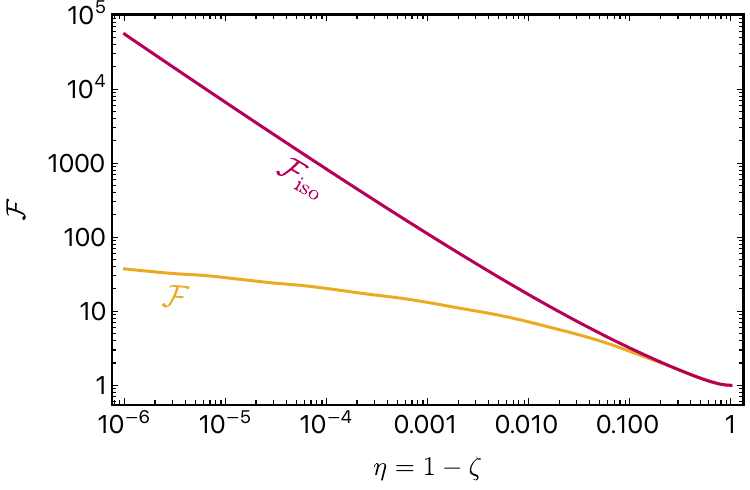}
    \caption{Anharmonic enhancement of the majoron energy density and isocurvature as a function of $\eta = 1-\zeta$}
    \label{fig:anharmonic}
\end{figure}
In the pre-inflationary scenario, inflation selects a nearly homogeneous initial value of the majoron field across our observable Universe, and the initial misalignment angle $\theta_i$ is therefore a free parameter. During the early radiation era, $H \gg m_\phi$ and the majoron is initially frozen due to Hubble friction. When the expansion rate drops below $H\sim m_\phi$, the majoron begins coherent oscillations around the minimum of its potential. The energy stored in these oscillations subsequently redshifts as nonrelativistic matter and gives the misalignment contribution to the majoron relic abundance. Thus, the relic density is controlled by the misalignment angle $\theta_i$.

The majoron energy density today is given by 
 \begin{align}
     \label{eq:shortinfomega}
    \Omega_{\phi} & =  \frac{ \tfrac{1}{2} \, m^{2}_{\phi} \, f_\phi^2 \, \theta_{\rm osc}^{2} \mathcal{F}(\zeta)}{\rho_{\rm crit}}
    \frac{ g_{*S,0} \, T_{0}^3}{g_{*S,{\rm osc}} \, T^3_{\rm osc}} \\
&      \simeq 0.2 \times \zeta^{2} \mathcal{F}(\zeta) \, \left(\frac{f_\phi}{10^{10} \, \GeV}\right)^{2} \left(\frac{m_{\phi}}{100 \, {\rm keV}}\right)^{1/2} \left(\frac{106.75}{g_{*,{\rm osc}}}\right)^{1/4} \, . \nonumber
 \end{align} 
Here $\theta_{\rm osc} \simeq 0.63 \, \theta_{i}$, and $\zeta \equiv \theta_{i}/\pi$. The critical density today is $\rho_{\rm crit} = 3 H^{2}_{0} M^{2}_{\rm pl}/(8\pi)$, $T_0$ is the CMB temperature today, and $T_{\rm osc}$ is the temperature at which the majoron starts to oscillate, determined by the condition $3 H(T_{\rm osc}) = m_{\phi}$. The number of relativistic (entropy) degrees of freedom is denoted by $g_*$ ($g_{*S}$). Finally, the factor $\mathcal{F}(\zeta)$ accounts for the anharmonic correction to the majoron energy density arising from the cosine potential in Eq.~\eqref{eq:majoron}. We compute $\mathcal{F}$ numerically by solving the majoron background equation of motion as $\zeta$ is varied, with the result shown in Fig.~\ref{fig:anharmonic}; further details are provided in Appendix~\ref{app:Fcorrection}. 

For $\zeta=\theta_i/\pi\ll1$, the majoron potential in Eq.~(\ref{eq:majoron}) is well described by its quadratic approximation near the minimum, thus the evolution of the majoron is essentially harmonic and $\mathcal{F}(\zeta)\simeq 1$. 
However, if $\theta_i\to\pi$ $(\zeta\to1)$, two effects contribute to the deviation from the harmonic approximation. The first comes from the delayed onset of oscillations when the majoron is near the top of the potential, which enhances the energy density compared to the harmonic approximation. The second comes from the fact that, during the initial stages of oscillation, the majoron is sensitive to the full cosine potential, and its energy density does not scale as matter, $\rho_\phi\propto a^{-3}$, until the oscillations become small enough for the harmonic approximation to apply.

\begin{figure}
    \centering
    \includegraphics[width = 0.9\linewidth]{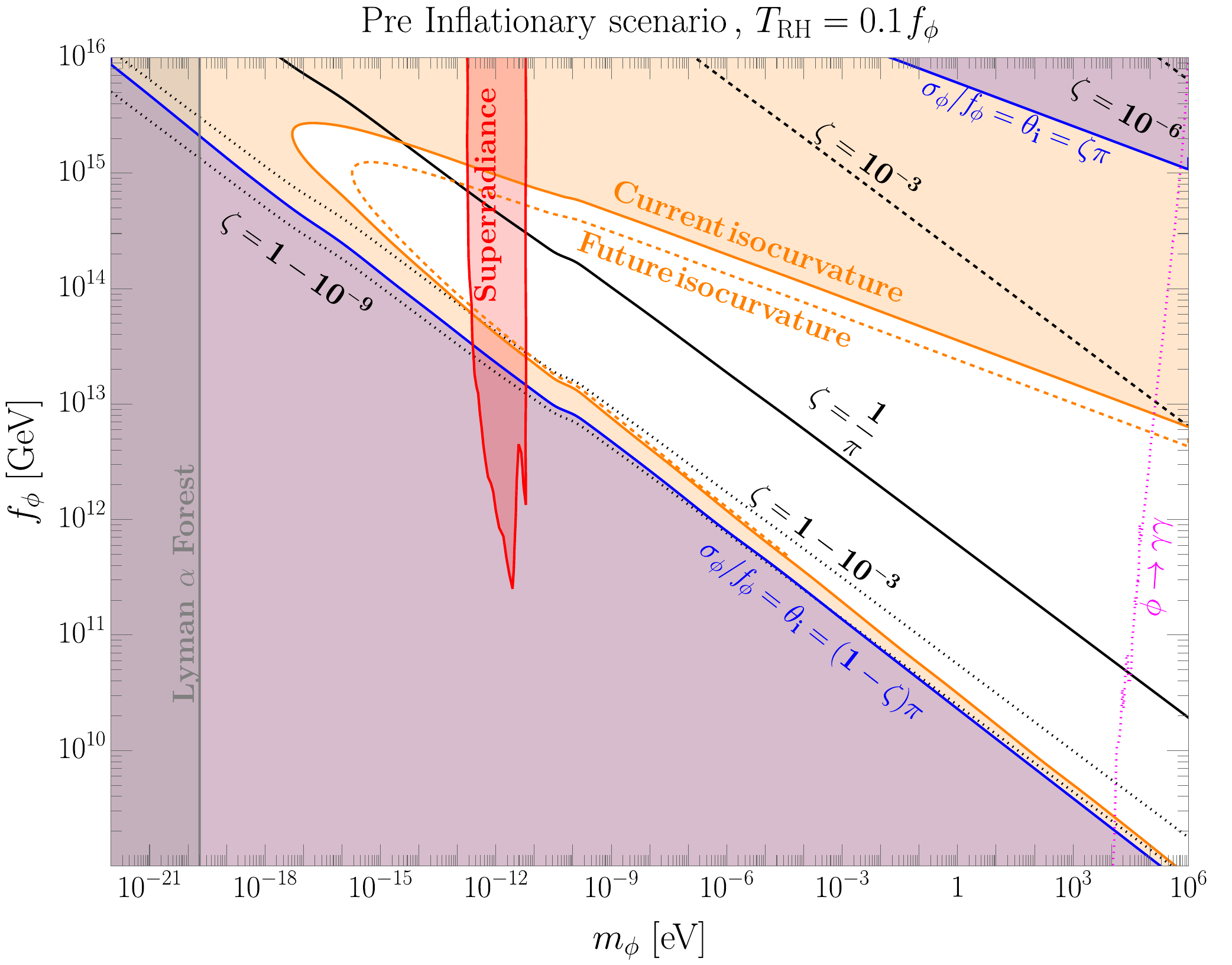}
    \caption{
    Majoron DM $m_\phi-f_\phi$ parameter space in the pre-inflationary scenario for the benchmark reheating temperature $T_{\rm RH}=0.1 f_\phi$, chosen to avoid thermal restoration of the $U(1)_L$ symmetry. At each point, the initial angle is chosen so that misalignment production (\ref{eq:shortinfomega}) accounts for all of the DM. 
    Black contours show the required initial angle: the solid contour denotes $\theta_i=1$, dashed contours denote small angle values $\theta_i=\zeta\pi$ with $\zeta\ll1$, and dotted contours denote hilltop values $\theta_i=(1-\eta)\pi$ with $\eta\ll1$. The blue region indicates where the required tuning of $\theta_i$ is smaller than the r.m.s inflationary fluctuation of the majoron (\ref{eq:constraintfluctuations1},\ref{eq:constraintfluctuations2}), and is therefore disfavored absent an additional mechanism. The orange region is excluded by current CMB isocurvature bounds, while the dashed orange contour shows expected future sensitivity. The red region is excluded by black hole superradiance constraints. The magenta line shows the boundary of the constraint from majoron decays to photons, evaluated using Eq.~\eqref{eq:tauagg} and maximal perturbative $R$ matrix elements. The gray shaded region indicates constraints from Lyman-$\alpha$ forest observations. Further details and references are provided in the main text. 
    }

    \label{fig:maxreheating}
\end{figure}

\begin{figure}[t]
\begin{center}
\includegraphics[width=0.495\linewidth]{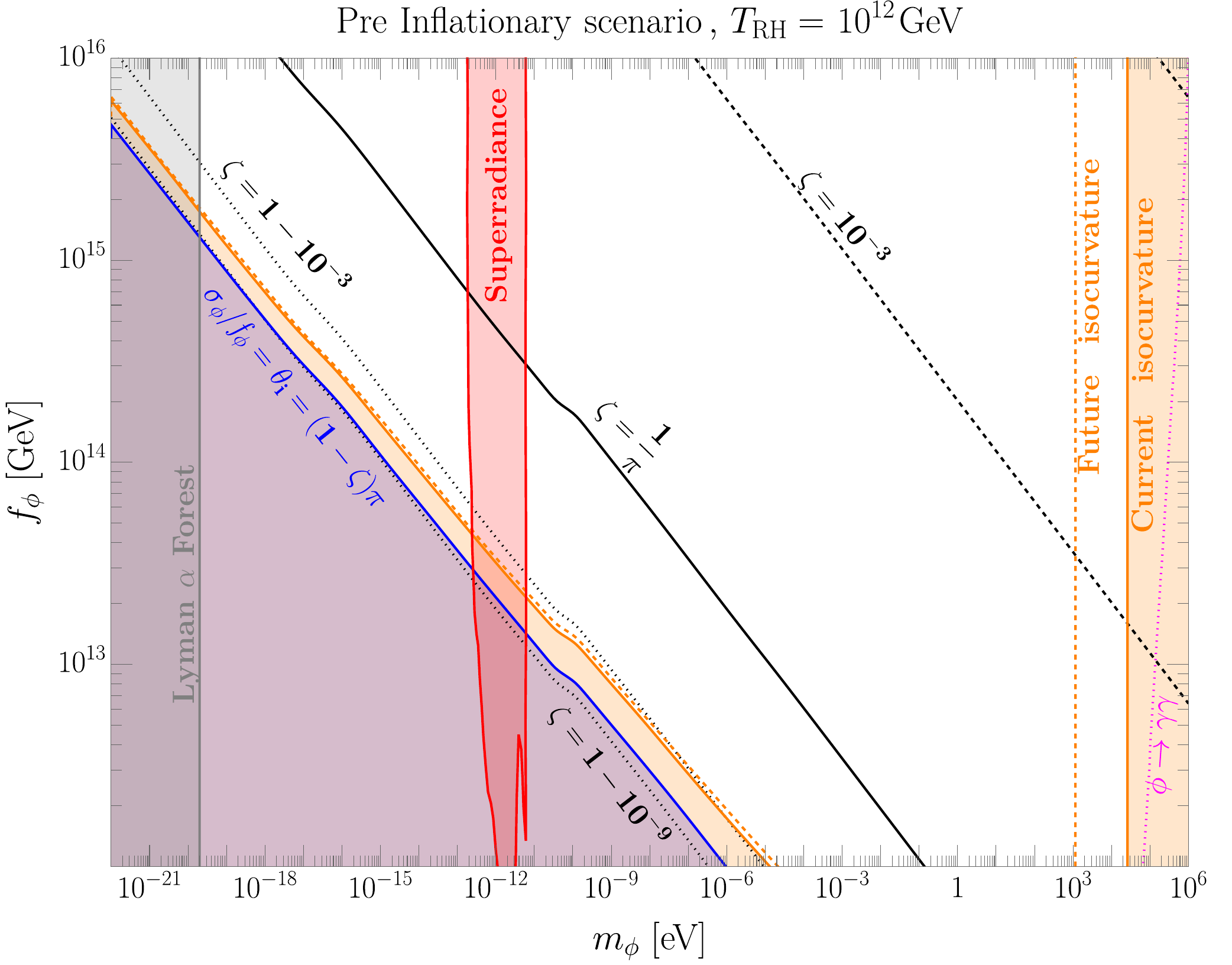}
\includegraphics[width=0.495\linewidth]{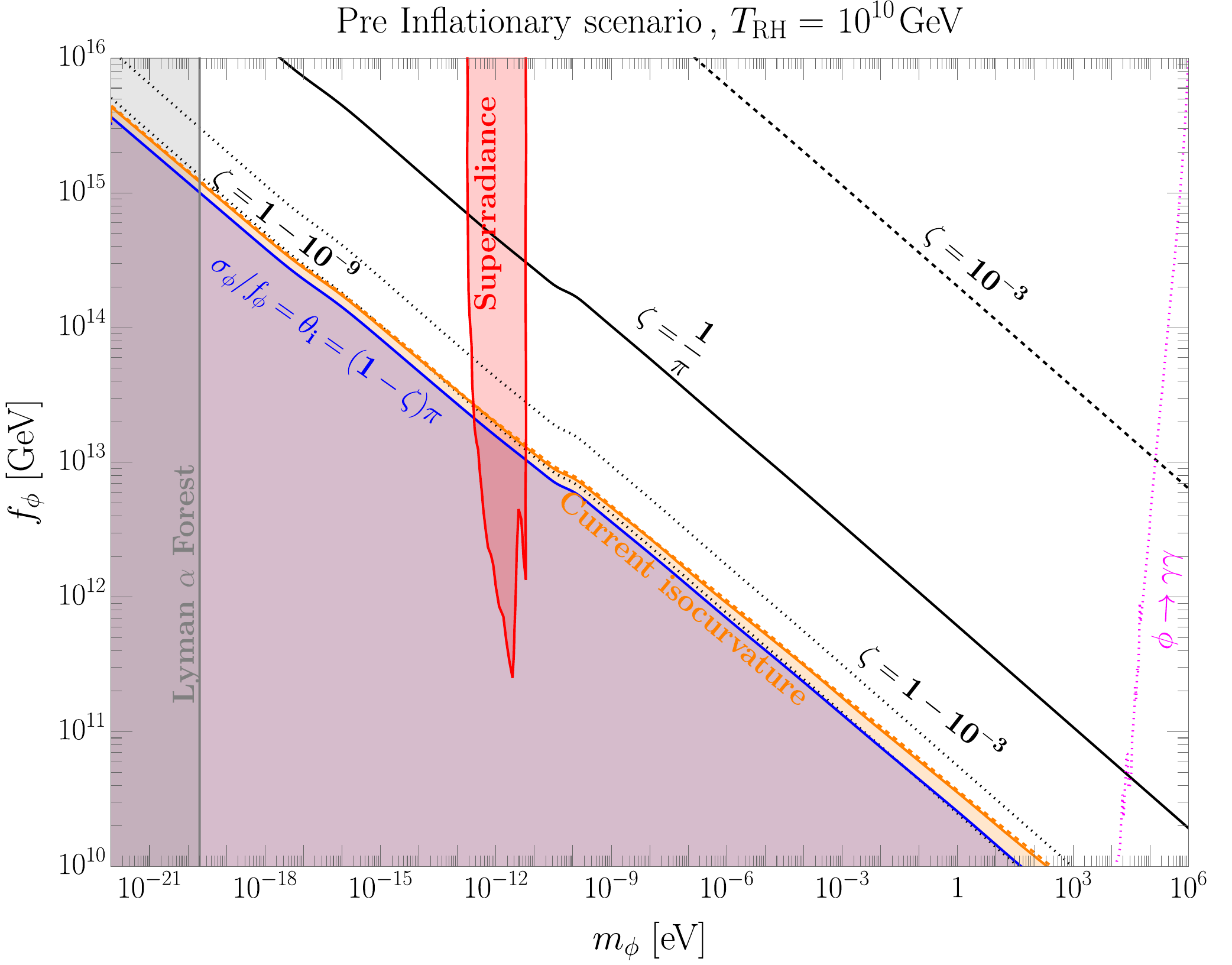}
\end{center}    
\caption{Same as Fig.~\ref{fig:maxreheating}, but for fixed $T_{\rm RH} = 10^{12} \, \GeV$ (left) and $T_{\rm RH} = 10^{10} \,\GeV$ (right). }
\label{fig:fixedTRH}
\end{figure}

The relic density estimate in Eq.~(\ref{eq:shortinfomega}) should be compared to the observed DM density $\Omega_{\rm DM} = 0.26$ \cite{Planck:2018vyg}. Figs.~\ref{fig:maxreheating} and~\ref{fig:fixedTRH} show the corresponding $m_\phi-f_\phi$ parameter space, with $\theta_i$ chosen at each point to reproduce the observed relic density. Isocontours indicate the required value of $\zeta\equiv\theta_i/\pi$. 
As is clear from Eq.~(\ref{eq:shortinfomega}), the required initial angle is correlated with the majoron mass and decay constant.
Larger values of $m_\phi$ and $f_\phi$ require smaller initial angles to produce the observed DM density, whereas smaller majoron masses and decay constants can require larger initial angles, approaching the hilltop regime where anharmonic effects enhance the abundance.

If there is no mechanism that selects a particular initial angle, it is natural to take $\theta_i$ to be uniformly sampled over the vauum manifold $-\pi<\theta_i\leq\pi$, i.e., the probability distribution is given by $p(\theta_i) = 1/(2\pi)$.
Both the small angle regime, $\zeta\ll 1$, where the field is initially close
to the minimum of the potential, and the hilltop regime,
$\eta\equiv 1-\zeta\ll 1$, where the field is initially close to the maximum,
correspond to special choices of initial conditions. We remain agnostic about the origin of the initial condition and treat $\theta_i$ as a free parameter. It is worth noting that such special choices of initial conditions could be selected, for example, by
additional dynamics or anthropic considerations, as has been discussed in the
axion context~\cite{Linde:1987bx,Wilczek:2004cr,Tegmark:2005dy,Hertzberg:2008wr,Takahashi:2019pqf,Co:2018mho,Co:2018phi,Co:2019jts}.
Absent such a dynamical or anthropic explanation, inflationary fluctuations of the majoron make extremely small values of $\zeta$ or $\eta$ increasingly unlikely. 
For a light spectator majoron satisfying $m_\phi \ll H_I$, and for the number of inflationary e-folds satisfying $N_e \ll H^{2}_{I}/m^2_\phi$, the rms fluctuation of the majoron field due to quantum fluctuations during inflation is given by~\cite{Starobinsky:1986fx,Starobinsky:1994bd,Graham:2018jyp,Takahashi:2018tdu}
\begin{align}
    \label{eq:vara}
    \sigma_{\phi} = \sqrt{N_e} \, \frac{H_{I}}{2\pi } \, .
\end{align}
Requiring that $\sigma_{\phi}$ is smaller than the distance between the $\phi_i$ and zero, and $\phi_i$ and $\pi f_\phi$, we obtain the respective conditions 
\begin{align}
    \label{eq:constraintfluctuations1}
    \frac{\sqrt{N_e}}{2\pi^{2}} \sqrt{\frac{4\pi^{3} g_{*,{\rm RH}}}{45}} \frac{T^{2}_{\rm RH}}{M_{\rm pl} f_{\phi}} &\lesssim \zeta 
    ~~~~\qquad\quad {\rm for~} \quad \zeta \rightarrow 0 \, ,\\
        \label{eq:constraintfluctuations2}
    \frac{\sqrt{N_e}}{2\pi^{2}} \sqrt{\frac{4\pi^{3} g_{*,{\rm RH}}}{45}} \frac{T^{2}_{\rm RH}}{M_{\rm pl} f_{\phi}} &\lesssim  (1-\zeta) ~~\quad {\rm for~} \quad \zeta \rightarrow 1 \, ,
\end{align}
Here we have assumed instantaneous reheating, which leads to the relation
\begin{equation}
\label{eq:inst-rh}
H_I\simeq \sqrt{\frac{4\pi^3 g_{*,{\rm RH}}}{45}}\,\frac{T_{\rm RH}^2}{M_{\rm pl}}.
\end{equation}
The constraint on $T_{\rm RH}$ will typically be more stringent if reheating is not instantaneous, since delayed reheating may imply a larger $H_I$ for a given $T_{\rm RH}$ and hence a larger $\sigma_\phi$ in Eq.~\eqref{eq:vara}.
The regions of majoron parameter space that do not satisfy the tuning criteria in Eqs.~(\ref{eq:constraintfluctuations1},\ref{eq:constraintfluctuations2}) 
are shown in Figs.~\ref{fig:maxreheating} and~\ref{fig:fixedTRH}. We assume $N_e=60$ and show results for 
 the benchmark $T_{\rm RH}=0.1 f_\phi$, corresponding to the maximal
reheating temperature we consider for the pre-inflationary scenario, and for the lower fixed values
$T_{\rm RH}=10^{12}\,\GeV$ and $10^{10}\,\GeV$.

In the above discussion and throughout this work, we have assumed the idealized scenario of instantaneous reheating. It is important to emphasize, however, that reheating is in general not instantaneous. In the standard perturbative picture, inflation is often followed by an extended period during which the inflaton oscillates about the minimum of its potential and its energy is gradually transferred to SM particles. For an approximately quadratic inflaton potential, this epoch is effectively matter dominated. During this period, the maximum temperature attained by the radiation bath can be parametrically larger than the reheating temperature~\cite{Giudice:2000ex}. Nevertheless, there exist reheating mechanisms in which the energy stored in the inflaton condensate is transferred to radiation on a timescale much shorter than a Hubble time, thereby approximately realizing the instantaneous reheating limit, as can occur, for example, in the instant preheating scenario explored in Ref.~\cite{Felder:1998vq}.

We note that freeze-in processes can provide an additional contribution to the majoron abundance, through reactions such as $N N\to \phi\phi$, $L N\to H\phi$, $H N\to L\phi$, and $L H\to N\phi$. For the hierarchy $m_{N_1}\lesssim T_{\rm RH}\ll m_{N_{2,3}}$, compatible with vanilla thermal leptogenesis, the heavier RHNs $N_{2,3}$ are not thermally populated, so freeze-in processes involving these states are negligible. While a detailed study of freeze-in production through $N_1$ is beyond the scope of this work, we have estimated the contribution from the corresponding $N_1$-mediated processes and find that it remains subdominant for $\lambda_1\lesssim {\cal O}(10^{-2})$. For larger $\lambda_1$, freeze-in production can become important and may give a substantial contribution to the majoron DM abundance for $m_\phi\gtrsim \mathcal{O}(10\,{\rm eV})$. For lighter majorons, this population would remain relativistic or semi-relativistic for longer and would act like a dark radiation or warm/hot relic component rather than cold DM.

\subsection{Inflationary fluctuations and isocurvature constraints}

As discussed above, a light majoron acquires quantum fluctuations during inflation with amplitude $\delta\phi \sim H_I/(2\pi)$. Since these fluctuations are uncorrelated with the inflaton fluctuations, spatial variations in the majoron abundance are not aligned with the adiabatic curvature perturbations, giving rise to DM isocurvature modes. The resulting isocurvature power spectrum is nearly scale invariant, with amplitude $A_{\cal S}$ given by 
(see, e.g., \cite{Kobayashi:2013nva,Allali:2025pja}) 
\begin{align}
    \label{eq:isocurvature}
    A_{\cal S} = \left(\frac{\partial \log \rho_{\phi}}{\partial \theta_i} \frac{H_I}{2 \pi f_\phi} \right)^{2} = 
    \left(\frac{H_{I}}{\pi^{2} f_{\phi} \, \zeta}\right)^{2}\mathcal{F}^{2}_{\rm iso}(\zeta) \, ,
\end{align}
where $\mathcal{F}^2_{\rm iso}(\zeta)$ represents a correction factor due to anharmonic effects. In the limit $\zeta\rightarrow 0$ the majoron potential (\ref{eq:majoron}) is quadratic to a good approximation, and $\mathcal{F}_{\rm iso} \rightarrow 1$. We numerically determine $\mathcal{F}_{\rm iso}$ by evolving the superhorizon perturbations as $\zeta$ is varied, as discussed in detail in  Appendix~\ref{app:Fcorrection}. The result is shown in Fig.~\ref{fig:anharmonic}. 
We note that $\mathcal{F}_{\rm iso}$ is related to $\mathcal{F}$ according to $\mathcal{F}_{\rm iso}(\zeta) = 1+\zeta F'(\zeta)/(2F(\zeta))$.

Current CMB observations constrain the isocurvature power at the pivot scale $k_{*} = 0.05~ {\rm Mpc}^{-1}$ to be $A_{\cal S} < 0.038 A_{s}$, where $A_{s} = 2.1\times 10^{-9}$ is the measured amplitude of the adiabatic curvature power spectrum~\cite{Planck:2018jri}. A future CMB-S4 experiment will improve this bound by a factor of $\sim 4$, corresponding to $A_{\cal S} < 0.009 A_{s}$ \cite{CMB-S4:2016ple}.

The isocurvature bound has important implications for the pre-inflationary majoron parameter space relevant to thermal leptogenesis.
In the pre-inflationary scenario, the thermal bath after inflation must not restore the $U(1)_L$ symmetry. We implement this requirement by taking $T_{\rm RH}\lesssim 0.1 f_\phi$ as a conservative benchmark.
On the other hand, thermal leptogenesis requires a sufficiently high reheating temperature to populate the heavy RHNs. We therefore translate the isocurvature constraint, which depends on $H_I/f_\phi$, into the
$m_\phi-f_\phi$ plane by assuming instantaneous reheating, Eq.~(\ref{eq:inst-rh}).
This assumption is conservative, since for fixed $T_{\rm RH}$, delayed reheating can correspond to a larger $H_I$, leading to larger majoron fluctuations and hence stronger isocurvature constraints.

Taking for concreteness a maximal reheating benchmark of $T_{\rm RH}=0.1 f_\phi$ and assuming instantaneous reheating, the largest isocurvature amplitude is then parametrically
\begin{align}
    \label{eq:SDMmax}
    A_{{\cal S},{\rm max}} &\simeq 5\times 10^{-11} \left(\frac{f_\phi}{10^{13} ~\GeV}\right)^{4} \left(\frac{m_{\phi}}{10\,{\rm keV}}\right)^{1/2},
\end{align}
where we have used the relic abundance condition to fix $\zeta$. For the parameter values used in Eq.~\eqref{eq:SDMmax}, this corresponds to the small angle regime, $\zeta\ll 1$, so the anharmonic factors satisfy $\mathcal{F}(\zeta)\simeq \mathcal{F}_{\rm iso}(\zeta)\simeq 1$. The full numerical analysis retains these factors and is required for large values of $\zeta$, where anharmonic effects become important.

The isocurvature constraints on the scenario with the largest reheating temperature is shown in Fig.~\ref{fig:maxreheating}. However in general, we will consider $ m_{N_{1}} \lesssim T_{\rm RH} \lesssim  0.1 \, f_{\phi}$. In Fig.~\ref{fig:fixedTRH} we show the allowed parameter space and associated constraints for $T_{\rm RH} = 10^{12}$ GeV (left) and $T_{\rm RH} = 10^{10}$ GeV (right). 
The isocurvature bound depends on the initial angle required to obtain the observed relic abundance via Eq.~(\ref{eq:isocurvature}). For larger $m_\phi$ and $f_\phi$, producing the observed abundance requires smaller initial angles, which enhances the isocurvature amplitude. For smaller $m_\phi$ and $f_\phi$, obtaining the correct abundance requires larger initial angles, eventually approaching the hilltop region of the potential. In this regime, the isocurvature amplitude is further enhanced by the anharmonic factor $\mathcal{F}_{\rm iso}(\zeta)$. Finally, the overall strength of the bound depends on the reheating temperature. Under the assumption of instantaneous reheating, larger $T_{\rm RH}$ implies a larger inflationary Hubble scale and therefore larger majoron fluctuations, while smaller $T_{\rm RH}$ weakens the constraint.

\subsection{Other probes of pre-inflationary majorons}
\label{sec:pre-inf-other}

Here we comment on several additional probes of majoron DM in our scenario, including indirect searches for astrophysical X-rays and soft $\gamma$-rays, black hole superradiance, Lyman-$\alpha$ forest observations, and CMB bounds on the tensor-to-scalar ratio.

For $m_\phi\lesssim \MeV$, the decay $\phi\to\gamma\gamma$ (see Eq.~(\ref{eq:tauagg})) can be probed by X-ray and soft $\gamma$-ray observations, which are displayed in Fig.~\ref{fig:maxreheating}. The INTEGRAL satellite has measured the Galactic soft $\gamma$-ray spectrum~\cite{Bouchet:2008rp,Bouchet:2011fn}, and Ref.~\cite{Laha:2020ivk} used these data to place limits on DM decays to photons. We also include constraints from NuSTAR observations of M31, which probe photon lines in the $10-100~{\rm keV}$ range~\cite{Ng:2019gch}. 
Our contours in Fig.~\ref{fig:maxreheating} assume neutrino Yukawa couplings near the perturbativity bound, leading to the associated lifetime estimate for  $\tau_{\phi\rightarrow \gamma \gamma}$ given in Eq.~(\ref{eq:tauagg}).

The presence of an ultralight boson such as the majoron can trigger a superradiant instability around rapidly spinning black holes, extracting angular momentum and spinning them down~\cite{Arvanitaki:2010sy,Arvanitaki:2009fg}. Superradiance is most efficient in black holes with a Schwarzschild radius comparable to the Compton wavelength of the ultralight boson. 
Superradiance becomes less effective for bosons with larger self-interactions. 
Measurements of spins of black holes in X-ray binaries lead to constraints on the mass and self interactions of majorons with mass $m_\phi\simeq 10^{-12}\,{\rm eV}$~\cite{Baryakhtar:2020gao,Hoof:2024quk,Witte:2024drg}. 
Measurements of spins of supermassive black holes (SMBHs) have been reported in the literature and have been used to constrain boson masses between $10^{-18}\,{\rm eV} - 10^{-20}\,{\rm eV}$~\cite{Stott:2018opm,Davoudiasl:2019nlo}. 
However, due to systematic uncertainties with measurements of spin, combined with the difficulty of modeling the impact of the complicated astrophysical environment on the growth of superradiant instability~\cite{Baryakhtar:2020gao}, we do not show these constraints in Figs.~\ref{fig:maxreheating} and~\ref{fig:fixedTRH}.
See also Ref.~\cite{Lambiase:2025twn} for a study of superradiance in scanarios with ultralight bosons coupled to neutrinos.

For sufficiently light majoron DM, the large de Broglie wavelength of the field suppresses the growth of structure below a characteristic scale, reducing power on dwarf galaxy scales. Lyman-$\alpha$ forest measurements, which probe the matter power spectrum on these small scales, therefore place a lower bound on the mass; we adopt the representative constraint $m_\phi \gtrsim 2\times 10^{-20}\,{\rm eV}$~\cite{Rogers:2020ltq,Irsic:2017yje,Kobayashi:2017jcf}.

The Planck + BICEP non-observation of CMB B-modes associated with tensor fluctuations produced during inflation leads to an upper bound on the inflationary Hubble scale, $H_I < 4.7 \times 10^{13}$ GeV~\cite{BICEP:2021xfz}. Assuming instantaneous reheating 
(see Eq.~(\ref{eq:inst-rh})), this translate to the bound $T_{\rm RH} < 5.5 \times 10^{15}$ GeV. We thus restrict to decay constants satisfying the rough bound $f_\phi \lesssim 10^{16}$ GeV in Figs.~\ref{fig:maxreheating} and \ref{fig:fixedTRH}.

\section{Majoron dark matter in the post-inflationary scenario}
\label{sec:post-inf}

We next consider majoron cosmology in the \textit{post-inflationary} scenario, in which lepton number is restored following inflation and is spontaneously broken only later as the Universe cools. This scenario can arise if inflationary fluctuations are large compared to the symmetry-breaking scale, $H_I/(2\pi)\gtrsim f_\phi$, or if the maximum temperature attained after inflation exceeds the symmetry-breaking scale. Assuming instantaneous reheating for simplicity, the latter thermal restoration condition is the stronger criterion, since $T_{\rm RH}\sim \sqrt{M_{\rm pl}  H_I}\gg H_I/(2\pi)$. We therefore take the post-inflationary regime to be characterized parametrically by $T_{\rm RH}\gtrsim f_\phi$.

In this scenario, the $U(1)_L$-charged complex scalar $\sigma$ is initially localized near the origin before subsequently relaxing onto the vacuum manifold, at which point the majoron angle is randomly selected in each causally disconnected Hubble patch. 
We therefore take the initial majoron angle to be uniformly distributed over $-\pi<\theta\leq\pi$, with probability density $p(\theta)=1/(2\pi)$.
As a result, the majoron DM abundance receives a misalignment contribution averaged over the initial angle, as well as additional contributions from the global string and domain-wall network.
In contrast to the pre-inflationary scenario, where the initial angle is a free parameter, the majoron relic density in the post-inflationary scenario is largely determined by $m_\phi$ and $f_\phi$.
Compatibility with standard thermal leptogenesis requires the reheating temperature to be high enough to thermally produce the lightest RHN. Combining this with the requirement that lepton number be restored after inflation, we consider the schematic hierarchy of scales $m_{N_1}\lesssim f_\phi\lesssim T_{\rm RH}$ for the post-inflationary scenario. 

The fact that lepton number is thermally restored in this scenario implies that the $U(1)_L$ charged scalar sector is thermally populated at high temperatures. We therefore estimate the thermally populated majoron contribution to the abundance, in addition to the nonthermal contributions discussed above.

\subsection{\textbf{Majoron abundance from misalignment, cosmic strings, and domain walls}}
\label{sec:strings}

We first consider the contribution to the majoron abundance from misalignment. 
In each causally disconnected patch, the local evolution is the same as in the pre-inflationary scenario, but with an initial angle $\theta_i$ drawn from the uniform distribution on the vacuum manifold. 
The spatially averaged misalignment contribution to the majoron abundance, including anharmonic effects, is then obtained by averaging Eq.~(\ref{eq:shortinfomega}) over this distribution: 
\begin{align}
    \label{eq:averageomega}
    \Omega_{\phi,{\rm mis}} &= \frac{\frac{1}{2}\pi^{2} (0.63)^{2} f^{2}_{\phi} m^{2}_{\phi}}{\rho_{\rm crit}} 
     \frac{g_{*S,0} \, T^3_{0}}{ g_{*S,{\rm osc}} \,T^3_{\rm osc}}
    \left(\int^{1}_{0} \zeta^{2} \mathcal{F}(\zeta) d\zeta\right)\nonumber\\
    &\simeq  0.2 \times \left(\frac{f_{\phi}}{10^{11} ~\GeV}\right)^{2} \left(\frac{m_{\phi}}{10 ~{\rm eV}}\right)^{1/2} \left(\frac{g_{*,{\rm osc}}}{106.75}\right)^{-1/4} \, .
\end{align}
As discussed above, $\zeta=\theta_i/\pi$, $T_{\rm osc}$ denotes the onset of oscillations, defined by $3H(T_{\rm osc})=m_\phi$, and we have used $\theta_{\rm osc}\simeq0.63\,\theta_i$. Additionally, $\mathcal{F}(\zeta)$ denotes the anharmonic correction factor; see Fig.~\ref{fig:anharmonic} and Appendix~\ref{app:Fcorrection}.

In the post-inflationary scenario, spontaneous breaking of $U(1)_L$ produces a network of global cosmic strings via the Kibble mechanism~\cite{Kibble:1976sj,Vilenkin:1984ib}. 
As the Universe expands, the string network evolves towards a scaling regime~\cite{Kibble:1976sj,Vilenkin:1984ib,Albrecht:1984xv,Bennett:1987vf,Allen:1990tv}, with a roughly constant number of strings per Hubble patch up to logarithmic corrections~\cite{Gorghetto:2018myk,Kawasaki:2018bzv}, while curved strings and loops lose energy by radiating majorons.
This provides an additional, nonthermal source of majoron DM beyond the misalignment contribution (\ref{eq:averageomega}). Furthermore, when the explicit symmetry-breaking potential later becomes dynamically important, domain walls form and attach to the strings, leading to the eventual collapse of the string-wall network and further majoron production.

The dynamics of such networks have been extensively studied in the context of axion strings~\cite{Davis:1986xc,Vilenkin:1982ks,Sikivie:1982qv,Harari:1987ht,Shellard:1987bv,Hagmann:1990mj,Battye:1993jv,Battye:1994au,Yamaguchi:1998gx,Hiramatsu:2010yu,Klaer:2017ond,Vaquero:2018tib,Gorghetto:2018myk,Kawasaki:2018bzv,Drew:2019mzc,Dine:2020pds,Gorghetto:2020qws,Buschmann:2021sdq,Drew:2022iqz,Drew:2023ptp,Kim:2024wku,Saikawa:2024bta,Kim:2024dtq,Buschmann:2024bfj,Benabou:2024msj}. While analytic arguments provide useful insight into their scaling behavior, a quantitative treatment of the string contribution to the majoron abundance generally requires numerical simulations. 
However, because of the large hierarchy between the string core scale, set by the inverse radial mode mass $m_r^{-1}\sim f_\phi^{-1}$, and the horizon scale $H^{-1}$, simulations of global string dynamics require extrapolation over many orders of magnitude. As a result, the string contribution to the majoron relic abundance remains subject to significant theoretical uncertainties.

A key quantity is the average number of strings per Hubble patch, $\xi(t)$. Numerical simulations of string networks suggest a logarithmic violation of the scaling behavior,
\begin{equation}
    \label{eq:scaling-deviation}
    \xi(t) \simeq 0.21 \log{\left(\frac{m_r}{H(t)}\right)},
\end{equation}
where the overall coefficient is taken from recent simulations~\cite{Kim:2024wku,Benabou:2024msj}.
The energy density of the strings is given by $\rho_s = \xi(t) \mu(t) /t^2$, where the effective string tension is $\mu(t) = \pi f_\phi^2 \log[m_r / (H(t) \sqrt{4\pi\xi(t)}) ]$.

Simulations support a power law parameterization of the instantaneous majoron emission spectrum, characterized by a spectral index $q$. Following Ref.~\cite{Kim:2024wku}, and specializing to the case $q=1$, the differential energy density of emitted majorons can be written, up to higher order corrections in $1/\log(m_r/H)$, as
\begin{equation}
    \label{eq:MajoronDiffEnergyDensity}
    \frac{\partial \rho_\phi}{\partial k}(t)
    \approx
    \frac{8H^2(t)\mu(t)\xi(t)}{k}
    \frac{\log\!\left(k/k_{\rm min}\right)}
    {\log\!\left(m_r/H(t)\right)} ,
    \qquad
    k_{\rm min}\leq k\leq k_{\rm max}.
\end{equation}
Here $k$ is the majoron momentum, $k_{\rm min}=x_0\sqrt{\xi(t)}\,H(t)$, with $x_0=10$ motivated by simulations~\cite{Kim:2024wku}, and $k_{\rm max}\sim\sqrt{m_r H(t)}$. The choice $q=1$ is supported by recent simulations~\cite{Benabou:2024msj} and analytic arguments~\cite{Dine:2020pds}, although some simulations find $q>1$; see, e.g., Ref.~\cite{Gorghetto:2020qws}.

Since the majoron number density produced at an earlier time $t_i$ is diluted by $(a_i/a)^3=(t_i/t)^{3/2}$ during radiation domination, the surviving number density is dominated by emission at late times. We therefore estimate the majoron abundance from strings near the time $t_*$, defined by $3H(t_*)\simeq m_\phi$, when the majoron mass becomes dynamically important and domain walls form. The number density of majorons emitted near $t_*$ is then given by
\begin{equation}
    \label{eq:numberdensity}
    n_\phi(t_*)=\int_{k_{\textrm{min}}}^{k_{\textrm{max}}} \frac{1}{k}\frac{\partial \rho_\phi}{\partial k}(t_*)dk \approx \frac{4\pi}{15}{f_\phi}^2 m_\phi \sqrt{\xi(t_*)} \, .
\end{equation}
The contribution to the present-day majoron density from strings is then
\begin{align}
    \label{eq:fstring}
    \Omega_{\phi,\rm{str}} & \simeq \frac{m_{\phi} \, n_{\phi}(t_{0})}{\, \rho_{\rm crit}} =  \frac{m_\phi \, n_{\phi}(t_{*})}{  \rho_\textrm{crit}}  \frac{g_{*S,0} \, T^3_{0}}{ g_{*S,*} \,T^3_{*}}\\
    &= 0.3\times \left(\frac{f_{\phi}}{10^{11} \, \GeV}\right)^{2} \left(\frac{m_{\phi}}{10 \, {\rm eV}}\right)^{1/2} \left(\frac{
    g_{*S,*}}{106.75}\right)^{-1/4}\left(\frac{\xi(t_{*})}{10}\right)^{1/2} \, . \nonumber
\end{align}
This is of the same order as the misalignment contribution in Eq.~(\ref{eq:averageomega}). Thus, for the majoron emission spectral index choice $q=1$
adopted here, strings provide only an $\mathcal{O}(1)$ enhancement to the majoron abundance. 
This estimate should be regarded as a benchmark, since it relies on extrapolating global string simulations to the physical hierarchy $\log(m_r/H)\gg10$ and is subject to statistical and systematic uncertainties. Moreover, simulations that find a steeper spectrum, $q>1$, imply emission dominated by infrared modes and can lead to a substantially larger majoron number density, potentially enhancing the string contribution by one to two orders of magnitude. 

After $t_{*}$ domain walls form and, for the $N_{\rm DW}=1$ case considered in here, the string-wall network collapses shortly afterwards.
Most of the energy of the network is converted into low momentum majorons.
Following Ref.~\cite{Benabou:2024msj}, we estimate the additional production from the collapse of the string-wall network by multiplying the early string emission contribution~(\ref{eq:fstring}) by the factor
\begin{equation}
\label{eq:RDW}
{\cal R}_{\rm str-dw} =
1+0.65+0.007
+4.9\left(\frac{\xi(t_*)}{13}\right)^{1/2}
+1.6\left(\frac{\xi(t_*)}{13}\right)^{-1/2}.
\end{equation}
The first three terms account for the pre-$t_*$ string contribution and the post-$t_*$ emission from subhorizon and superhorizon strings, while the last two terms account for majoron production from subhorizon and superhorizon domain walls, respectively.
For $\xi_*=13$, this gives ${\cal R}_{\rm str-dw}\simeq 8.2$. 
We emphasize, however, that the size of the string-domain wall network contribution remains uncertain. Earlier simulations
found a domain wall contribution comparable to the string contribution, see, e.g., Ref.~\cite{Hiramatsu:2012gg}, and the domain wall estimate may be reduced by order one factors if the emitted axions are semi-relativistic or if the wall configurations radiate less efficiently.

\begin{figure}
    \centering
    \includegraphics[width = 0.9\linewidth]{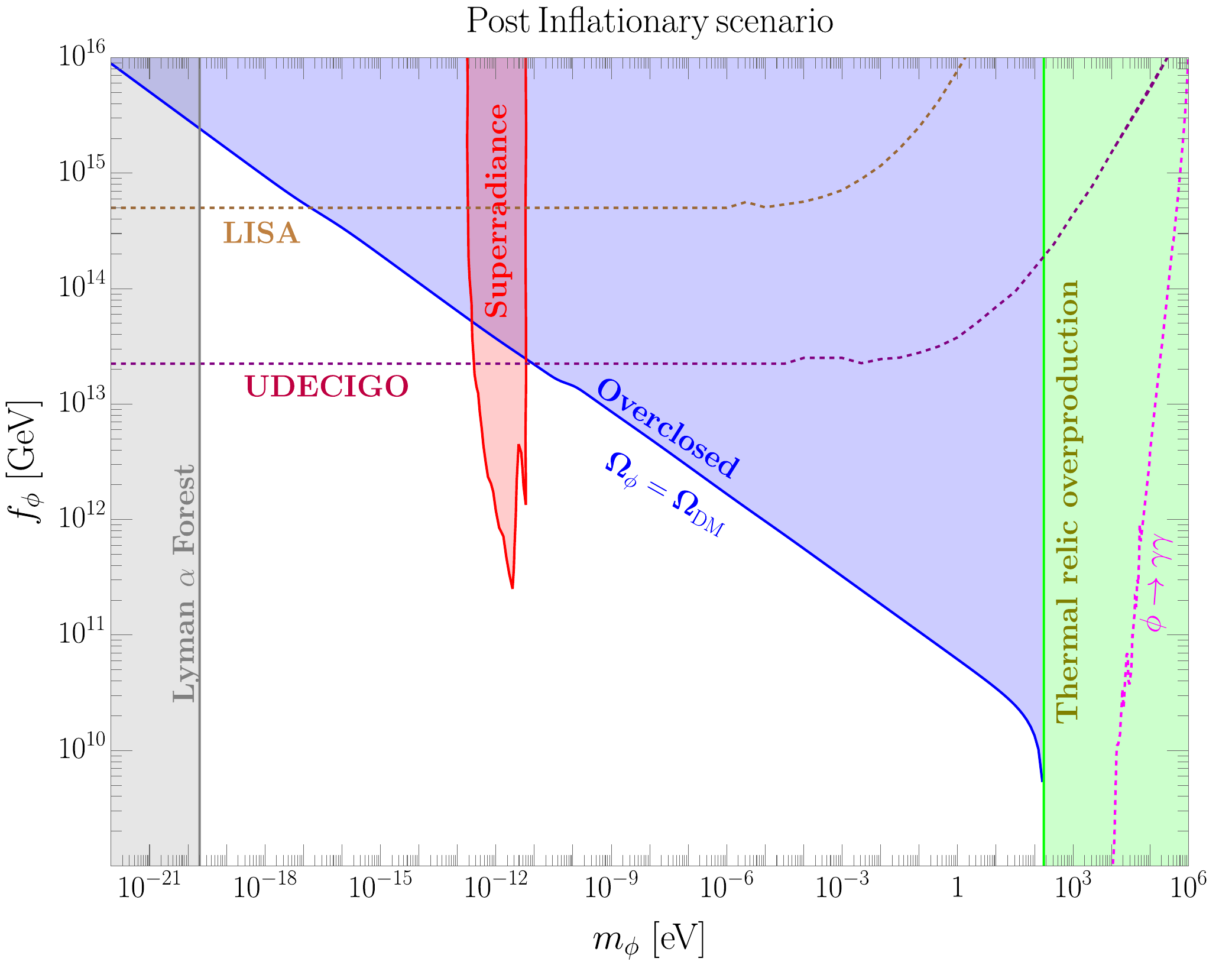}
    \caption{Majoron DM $m_\phi-f_\phi$ parameter space in the post-inflationary scenario. The majoron relic abundance is estimated by accounting for non-thermal production from misalignment, cosmic strings, and domain walls, as well as thermal production. Along the blue line, the total majoron abundance matches the observed DM abundance. 
    In the blue shaded region above the blue line, majorons are overproduced relative to the observed DM density, while in the white region below the line they are underproduced. The green shaded region indicates where the thermal relic abundance alone exceeds the observed DM abundance. The red shaded region is excluded by black hole superradiance constraints, while the region to the right of the magenta line is probed by majoron decays to photons. The brown dashed and violet dotted curves show the projected reaches of LISA and UDECIGO, respectively; above these curves, the stochastic gravitational wave background from the cosmic strings can potentially be probed. The gray shaded region indicates constraints from Lyman-$\alpha$ forest observations. Further details and references are provided in the main text. }
    \label{fig:postinflationary}
\end{figure}

We estimate the total majoron abundance to be $\Omega_{\phi,{\rm tot}} = \Omega_{\phi,{\rm mis}}+\Omega_{\phi,{\rm str}} {\cal R_{\rm str-dw}}+\Omega_{\phi,\rm th}$, the sum of the misalignment contribution in Eq.~(\ref{eq:averageomega}), the pre-$t_*$ string contribution in Eq.~(\ref{eq:fstring}) multiplied by the post-$t_*$ enhancement factor ${\cal R}_{\rm str-dw}$ in Eq.~(\ref{eq:RDW}), and the thermally populated majoron contribution (\ref{eq:thermalomega}), to be discussed below in Section~\ref{sec:therm}. Fig.~\ref{fig:postinflationary} shows the corresponding $\Omega_{\phi,{\rm tot}}=\Omega_{\rm DM}$ contour. 

In our discussion above we have assumed the minimal $N_{\rm DW}=1$ case (see Eq.~(\ref{eq:majoron})), for which the resulting string-wall network is unstable. Instead, for $N_{\rm DW}>1$, the string-wall network is stable if the degenerate vacua are exact. Since the domain wall energy density redshifts more slowly than radiation, such a network would eventually dominate the Universe, leading to the standard cosmological domain wall problem~\cite{Zeldovich:1974uw,Sikivie:1982qv}. This problem can be avoided by introducing an additional bias term that explicitly lifts the $Z_{N_{\rm DW}}$ degeneracy and causes the network to annihilate~\cite{Sikivie:1982qv,Gelmini:1988sf,Larsson:1996sp,Hiramatsu:2010yn,Correia:2014kqa,Correia:2018tty}. In that case, however, the cosmology becomes sensitive to the size and structure of the bias term. Moreover, if the annihilation occurs at $H\ll m_\phi$, the string-wall network can constitute a larger fraction of the total energy density before it decays, potentially enhancing both nonthermal majoron production and the associated gravitational wave signal (to be discussed below)~\cite{Hiramatsu:2013qaa,Saikawa:2017hiv,Hong:2025piv, Notari:2025kqq}. In this sense, focusing on $N_{\rm DW}=1$ provides a minimal and conservative benchmark.

\subsection{\textbf{Thermal production of majorons}}
\label{sec:therm}

In the post-inflationary scenario, thermal restoration of lepton number for $T_{\rm RH}>f_\phi$ requires the $U(1)_L$ complex scalar $\sigma$ to be sufficiently coupled to the primordial plasma. Consequently, the degrees of freedom of this scalar field are thermally populated before the phase transition, and after symmetry breaking the majoron inherits a thermal population. 
We therefore expect a thermal relic contribution to the majoron abundance in this scenario, in addition to nonthermal production from misalignment, strings, and domain walls.
The cosmological impact of this thermal population of majorons depends on their mass.  

If the majorons are heavy enough, $m_{\phi} \gtrsim  10^{-4} \, \rm{eV}$, they will become non-relativistic by today, with energy density
\begin{align}
    \label{eq:thermal}
    \rho_{\phi,{\rm th}} = \frac{\zeta(3)}{\pi^{2}} \, T^{3}_{\phi,0} \, m_{\phi} \, , 
\end{align}
where $T_{\phi,0}$ is the majoron temperature today. After decoupling, the majoron temperature redshifts as $T_\phi\propto a^{-1}$,
while the photon bath is reheated by subsequent entropy transfer from SM species.
Assuming standard cosmology, the majoron temperature today can be written as 
$T_{\phi,0} = T_{\gamma,0} \left(g_{*S,0}/g_{*S,{\rm dec}}\right)^{1/3} \simeq 0.33 \, T_{\gamma,0}$,
where $g_{*S,0} = 3.91$ and we take $g_{*S,{\rm dec}} = 106.75$ since the majoron decouples at very high temperatures. 
This corresponding majoron DM density today is given by
\begin{align}
    \label{eq:thermalomega}
    \Omega_{\phi,{\rm th}} = 1.5 \times 10^{-3}\left(\frac{m_{\phi}}{\rm{eV}}\right).
\end{align}
Requiring $\Omega_{\phi,{\rm th}} \leq \Omega_{\rm DM} = 0.26$, we obtain  an upper bound on the majoron mass,
\begin{align}
    \label{eq:massupperboundthermal}
    m_{\phi} < 170 \, {\rm eV}  \, .
\end{align}

Lighter majorons with mass $m_{\phi} \lesssim 10 \, {\rm eV}$ will contribute to dark radiation around the epoch of matter-radiation equality. The contribution of majorons to the effective number of neutrino species is 
\begin{align}
    \label{eq:neff}
    \Delta N_{\rm eff} \simeq 0.027 \, .
\end{align}
A contribution of this size may be within reach of future CMB experiments~\cite{CMB-S4:2016ple}.

We note that in the post-inflationary scenario, the thermal relic population of majorons is expected to dominate over any subsequent freeze-in production. We therefore neglect freeze-in contributions to the majoron abundance in this case.

\subsection{\textbf{Gravitational waves from cosmic strings}}
\label{GWIsoPostInf}

In the post-inflationary scenario, global cosmic strings also radiate gravitational waves before the string-wall network annihilates, producing a stochastic background that is potentially detectable in future gravitational wave observatories~\cite{Vilenkin:1981bx,Vachaspati:1984gt,Allen:1991bk,Battye:1993jv,Battye:1994au,Chang:2019mza,Servant:2023mwt,Fu:2023nrn}. 
As discussed above, after formation, the string network evolves toward a scaling regime in which the number of strings per Hubble volume is approximately constant, up to logarithmic corrections characteristic of global strings. Energy loss from the network is dominated by majoron radiation, while a subdominant fraction is emitted as gravitational waves, both primarily sourced by oscillating loops formed through the intercommutation of long strings.
The resulting spectrum is approximately scale invariant over the range of frequencies corresponding to the scaling regime, with an amplitude controlled by the string tension $\mu$ and suppressed relative to majoron radiation by the parametric factor $f_\phi^2/M_{\rm pl}^2$.
Following Refs.~\cite{Cui:2017ufi,Cui:2018rwi,Gouttenoire:2019kij} the gravitational wave spectrum can be written as
 \begin{equation}
     \label{eq:GW31}
     \begin{split}
         &\Omega_{\rm GW}(f)=\frac{1}{\rho_{\rm crit}}\frac{\mathcal{F}_\alpha}{\sum_{p=1}^{\infty} p^{-n}}  \frac{F_\alpha}{\alpha}\sum_{k=1}^{\infty}  \frac{2k}{f } \int_{t_F}^{t_{dec}}  d\tilde{t}\left[  C_{\textrm{eff}} (t_i^{(k)})\frac{1}{{(t_i^{(k)}})^4} {\left(\frac{a(t_i^{(k)})}{a(\tilde{t})}\right)}^3 \frac{\Gamma k^{-n} G\mu^2}{\alpha +\Gamma G\mu +\kappa}                 \right.\\
         &\left.  \hspace{220pt}   {\left(\frac{a(\tilde{t})}{a(t_0)}\right)}^5 \Theta(\tilde{t}-t_i^{(k)})\Theta(t_i^{(k)}-t_F)             \right] \, ,
     \end{split}
 \end{equation}
where $f$ is the frequency of the observed gravitational wave, 
$k$ denotes the normal mode of the oscillating loop, 
$C_{\rm eff} \sim {\cal O}(1)$ is the loop formation efficiency, 
$\mathcal{F}_\alpha$ is the loop size distribution at formation with $\alpha$ the fractional length of the loop relative to the horizon size, 
$F_\alpha$ is the fraction of energy stored in loops that can be radiated away as gravitational waves or majorons,
and $n = 4/3$ assuming radiation is dominanted by cusps in oscillating loops. We take $\alpha = 0.1$, with $F_\alpha = 0.1$ and $\mathcal{F}_\alpha = 0.1$ following Refs.~\cite{Chang:2021afa} and fix $C_{\rm eff} =1$~\cite{Chang:2021afa,Gouttenoire:2019kij}.
Furthermore, $t_i^{(k)}$ is the time at which a loop oscillating in the $k^{\textrm{th}}$ mode and emitting gravitational waves at time $\tilde{t}$ with frequency $\tilde{f}=[a(t_0)/a(\tilde{t})]\,f$ was created,
\begin{equation}
\label{eq:t-i-k}
t_i^{(k)}(\tilde{t},f)\simeq\frac{1}{\alpha+ \Gamma G \mu + \kappa} 
\left[\,\tilde{l}+\Gamma G \mu \,\tilde{t} + \kappa\,\tilde{t}\,\right] \, ,
\end{equation}
where $\tilde{l}(f,k,\tilde{t})= (a(\tilde{t})/a(t_0))(2k/f)$ is the length of the corresponding loop. 
Additionally, the formation time $t_F$ of the string network is defined by the condition $\sqrt{\rho_{\textrm{tot}}(t_F)}=\mu$ where $\rho_{\textrm{tot}}$ is the total energy density of the universe~\cite{Gouttenoire:2019kij}, while $t_{\rm dec}$ corresponds to the time of collapse of the string-wall network, determined by the condition $3H(t_{\textrm{dec}})= m_{\phi}$~\cite{Servant:2023mwt}. Finally, $\rho_{\rm crit}$ is the critical density, $\Gamma \simeq 50$ is a constant which is proportional to the rate at which the oscillating loops radiate energy into gravitational waves~\cite{Vachaspati:1984gt,Chang:2021afa,Gouttenoire:2019kij,Blanco-Pillado:2013qja,Blanco-Pillado:2017oxo}, 
$\kappa = \Gamma_\phi /(2 \pi \log[f_\phi \, \tilde t\,] \,)$ with $\Gamma_\phi\simeq65$ represents the rate at which oscillating loops radiate energy to majorons~\cite{Vilenkin:1986ku,Chang:2021afa,Gouttenoire:2019kij}
and $G =1/M_{\rm pl}^2$ is Newton's constant.  

\begin{figure}
    \centering
    \includegraphics[width = 0.9\linewidth]{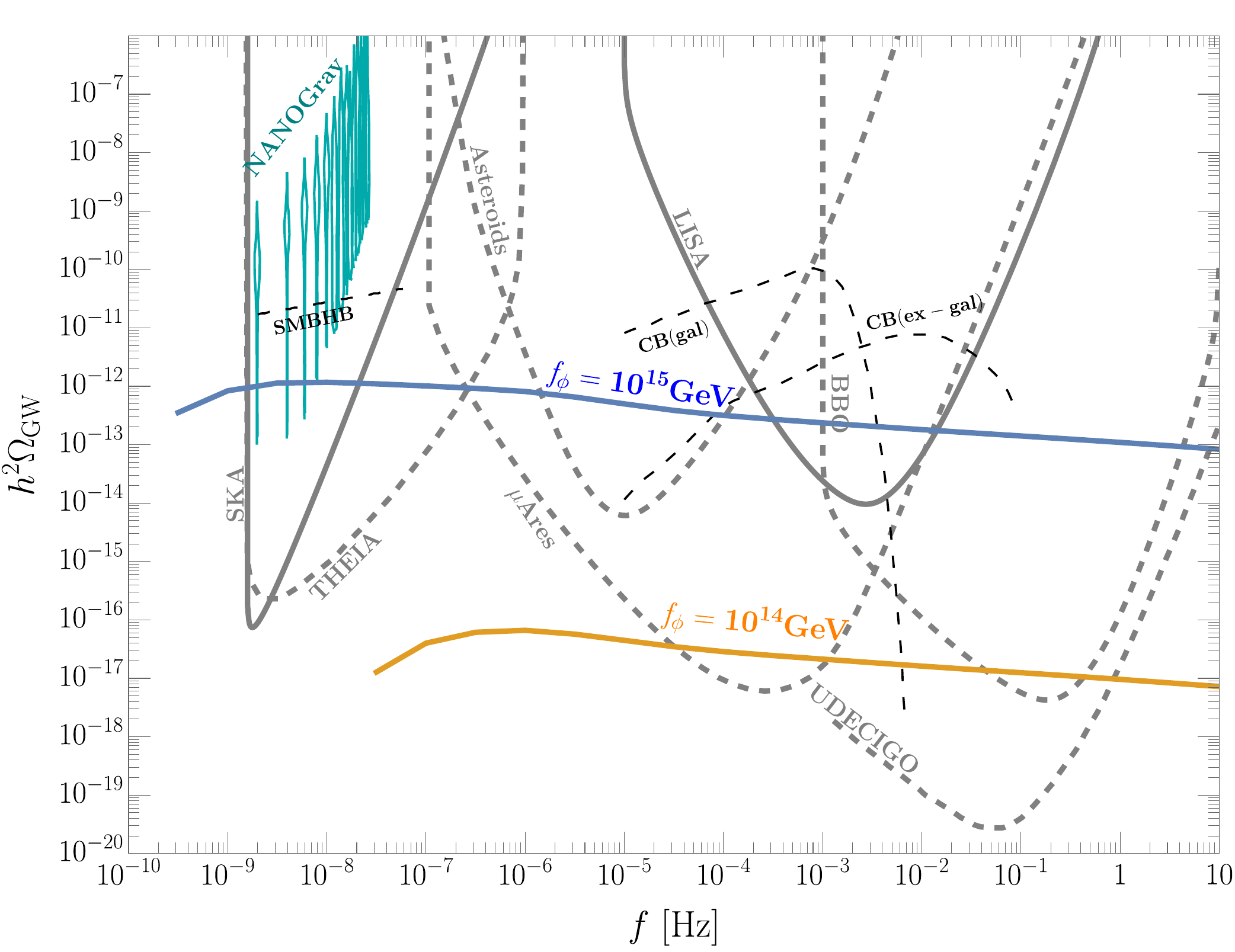}
    \caption{
    Gravitational wave spectra from cosmic strings in the post-inflationary scenario for $f_{\phi} = 10^{14} \,\GeV$ (orange) and $f_{\phi} = 10^{15}\, \GeV$ (blue). Sensitivities from representative current and future gravitational wave experiments are overlaid. Further details are provided in the main text.} 
    \label{fig:GW}
\end{figure}

In Fig.~\ref{fig:GW}, we compare the predicted gravitational wave spectra for $f_\phi = 10^{14}$ GeV and $f_\phi = 10^{15}$ with power-law integrated sensitivity (PLIS) curves corresponding to a SNR threshold of 1 for several proposed observatories. These include the SKA PTA~\cite{Janssen:2014dka}; the space-based interferometers LISA~\cite{LISA:2017pwj}, BBO~\cite{Crowder:2005nr}, Ultimate-DECIGO~\cite{Braglia:2021fxn}, and $\mu$Ares~\cite{Sesana:2019vho}; asteroid laser ranging~\cite{Fedderke:2021kuy}; and future astrometric measurements~\cite{Moore:2017ity}, as envisioned for THEIA~\cite{2018FrASS...5...11V}. The sensitivity curves are taken from Refs.~\cite{Schmitz:2020syl,Janssen:2014dka,Caprini:2019pxz,Crowder:2005nr,Sesana:2019vho,Fedderke:2021kuy,Garcia-Bellido:2021zgu,Braglia:2021fxn}. We also show the NANOGrav 15-year stochastic gravitational wave spectrum~\cite{NANOGrav:2023hvm}, together with estimates of astrophysical foregrounds from supermassive black hole binaries (SMBHBs)~\cite{NANOGrav:2023gor}, galactic compact binaries~\cite{Cornish:2017vip}, and extragalactic compact binaries~\cite{Farmer:2003pa}. As shown in the figure, a range of proposed experiments could probe the predicted stochastic gravitational wave background for $f_\phi \gtrsim 10^{14}$ GeV. 

In Fig.~\ref{fig:postinflationary}, we translate the gravitational wave sensitivities of LISA and UDECIGO of Fig.~\ref{fig:GW} into projected reach in the majoron parameter space. For each point in the $m_\phi-f_\phi$ plane, we compute the resulting stochastic gravitational wave spectrum and indicate parameter values where the signal intersects (or is tangential to) the corresponding PLIS curve. We see that a significant portion of the majoron DM parameter space may be accessible to future gravitational wave observatories. 

For $N_{\rm DW}>1$, an additional bias can make the otherwise stable wall network
annihilate at $H \ll m_\phi$. In this case the walls may survive
long enough to enter a scaling regime, or even become a substantial fraction
of the total energy density, leading to enhanced majoron production and a
gravitational wave spectrum with a characteristic peak set by the horizon scale
at annihilation. This should be contrasted with the minimal $N_{\rm DW}=1$
case considered in this work, in which the string-wall network is unstable shortly after wall formation
and therefore does not undergo a prolonged metastable wall evolution.

\subsection{Other probes of post-inflationary majorons}

We now comment on several additional potential probes of the post-inflationary majoron. First, as in the pre-inflationary scenario, there are constraints from X-ray and soft $\gamma$-ray observations, black hole superradiance bounds, and Lyman-$\alpha$ forest observations. 
These are shown in Fig.~\ref{fig:postinflationary}. See Section~\ref{sec:pre-inf-other} for further details.

In the post-inflationary scenario, majoron production from strings and domain walls generally yields a population with nonzero momenta, so sufficiently light majorons may exhibit free streaming suppression of small-scale structure in addition to the usual suppression from the wave nature of ultralight DM. For relativistically produced wave DM, the free streaming scale can be logarithmically enhanced relative to the cold case, leading to somewhat stronger mass bounds. Ref.~\cite{Liu:2024pjg} finds a representative lower bound of order $m_\phi\gtrsim 5\times10^{-19}\,{\rm eV}$ in an analogous setting of axion/ALP fuzzy DM, and we expect that similar constraints apply to our post-inflationary majoron scenario. 

Post-inflationary majoron production can also generate white-noise isocurvature, since the initial field value and defect network are uncorrelated across causally disconnected regions, as in generic post-inflationary ALP DM scenarios~\cite{Feix:2019lpo,Feix:2020txt,Chathirathas:2025aan}. 
The power peaks near the causal scale at the onset of oscillations/string-wall annihilation, making this effect most relevant for very light majorons near the fuzzy DM regime and for small-scale probes such as Lyman-$\alpha$~\cite{Irsic:2019iff}, ultra-faint dwarf heating~\cite{Graham:2024hah}, and future 21-cm observations~\cite{Irsic:2019iff}. We note that current translated ALP bounds are not competitive with the free streaming constraints discussed above.

\section{Conclusions}
\label{sec:Conclusions}

We have studied majoron DM in a high-scale seesaw framework with spontaneously broken lepton number. This framework is compatible with high-scale thermal leptogenesis, consistent with the Davidson-Ibarra bound. In this setting, the same lepton number breaking dynamics that generate heavy Majorana masses for the RHNs also give rise to the majoron, a light pseudo-Nambu-Goldstone boson that can be cosmologically long-lived and constitute the DM of the Universe. We considered a minimal majoron potential with unit domain wall number and analyzed the resulting cosmology in both pre-inflationary and post-inflationary histories of lepton number breaking.

In the pre-inflationary scenario, lepton number is broken during inflation and remains broken afterward. The majoron field then takes a single coherent initial value throughout the observable Universe, and the relic abundance from misalignment is governed by the initial angle. Since this angle is a free parameter, the observed DM abundance can be obtained over a broad region of the $m_\phi-f_\phi$ parameter space. The required initial angle is correlated with the majoron mass and decay constant, with some regions requiring the field to start close to the minimum and others near the hilltop regime, where anharmonic effects become important. Inflationary fluctuations of the majoron lead to isocurvature perturbations, which are bounded by precision observations of the CMB. The isocurvature signal tends to be enhanced for larger reheating temperatures, smaller initial majoron angles, and in the hilltop regime where anharmonic effects are important. We find that isocurvature already imposes important constraints on the pre-inflationary scenario, while leaving viable majoron DM parameter space that can be further probed by future CMB observations.

In the post-inflationary scenario, lepton number is restored after inflation and breaks again during the subsequent thermal history. In this case, the initial majoron angle varies between causally disconnected patches, and the relic abundance is largely fixed once the majoron mass and decay constant are specified. The resulting majoron abundance receives contributions from misalignment, cosmic string radiation, string-domain wall network collapse, as well as a thermal population. This leads to a relic density target in the $m_\phi-f_\phi$ plane. For sufficiently large lepton number breaking scales, $f_\phi \gtrsim 10^{14}$ GeV, the associated cosmic string network can also source a stochastic gravitational wave background, providing an additional probe of the post-inflationary scenario in future gravitational wave experiments.

We also considered several other complementary probes of majoron DM. For ultralight majorons in the fuzzy DM regime, $m_\phi \lesssim 10^{-20}$ eV, the matter power spectrum is suppressed on small scales, leading to constraints from Lyman-$\alpha$ forest observations. Black hole superradiance provides sensitivity to majoron masses near $m_\phi\sim 10^{-12}$ eV. At higher majoron masses near the MeV scale, astrophysical searches for X-ray and soft $\gamma$-rays from $\phi\to\gamma\gamma$ decays provide another complementary probe. Together, these observations test distinct regions of the parameter space and illustrate the broad phenomenology of majoron DM.

\section*{Note added}

As this work was being completed, Refs.~\cite{Akita:2026gzk,deGiorgi:2026jqn} appeared. While these studies overlap with ours in considering majoron DM in seesaw-motivated frameworks, the analyses are complementary. Our work examines inflationary fluctuations and CMB isocurvature constraints in the pre-inflationary scenario, as well as thermal majoron populations and stochastic gravitational waves from  cosmic strings in the post-inflationary scenario. We also consider majoron masses extending down to the fuzzy DM regime. By contrast, Refs.~\cite{Akita:2026gzk,deGiorgi:2026jqn} provide detailed studies of freeze-in production, which is not the primary focus here. A preliminary version of our work was presented in Ref.~\cite{SDSUSY25}.

\acknowledgments
The work of BB and SD is supported by the U.S. Department of Energy under Grant No. DE–SC0007914. 
The work of AG is supported by the GRASP Initiative at Harvard University.

\appendix

\section{Anharmonic Corrections}
\label{app:Fcorrection}
\subsection{Majoron relic abundance}

The majoron relic abundance due to misalignment production is given in Eq.~(\ref{eq:shortinfomega}), including the factor 
$\mathcal{F}(\zeta)$ that accounts for anharmonic effects. 
In this appendix we describe our numerical approach to estimating $\mathcal{F}(\zeta)$.
The equation of motion for the majoron background, $\theta_0(t) = \phi_0(t)/f_\phi$, in the radiation-dominated epoch ($H=1/(2t)$), expressed in terms of dimensionless time parameter $y=m_{\phi} \, t$, is given by
\begin{equation}
    \label{eq:Anharmonic1}
    \frac{d^2 \theta_0}{dy^2}+ \frac{3}{2y}\frac{d\theta_0}{dy}+\sin{(\theta_0(y))}=0 \, .
\end{equation}
We numerically solve Eq.~\eqref{eq:Anharmonic1} in the time interval $y\in[y_i=10^{-3},y_f=10^{3}]$ with the initial conditions $\theta_0(y_i) = \theta_i =\zeta\pi$ and $d\theta_0/dy\big|_{y=y_i}=0$.

At late times, $y\gg1$, the expansion of the Universe has redshifted the majoron background $\theta_0(y)$ to relatively small values such that the harmonic approximation is valid. We therefore approximate the late time energy density as
\begin{equation}
    \label{eq:Anharmonic2}
    \rho_{\phi}(y,\zeta)=\frac{1}{2}{m_{\phi}}^2 {f_{\phi}}^2\left[{\left(\frac{d\theta_0(y)}{dy}\right)}^2+(\theta_0(y))^2\right]\hspace{5pt}.
\end{equation}
The anharmonic factor $\mathcal{F}(\zeta)$ is then defined as
\begin{equation}
    \mathcal{F}(\zeta) = \frac{\rho_\phi(y_f,\zeta)/\zeta^2}{\rho_\phi(y_f,\zeta_c)/\zeta_c^2} \, ,
\end{equation}
where $\zeta_{c} \ll 1$ is a reference value deep in the harmonic regime. In practice, we take $\zeta_{c} = 10^{-6}$.

\subsection{Majoron isocurvature perturbations}

The amplitude of the isocurvature power spectrum (\ref{eq:isocurvature}) depends on the anharmonic factor $\mathcal{F}_{\rm iso}(\zeta)$. We determine $\mathcal{F}_{\textrm{iso}}(\zeta)$ numerically using method described below. 

To compute the isocurvature power spectrum, we consider perturbations of the majoron about its homogeneous, time-dependent background, $\theta(t,\mathbf{x})=\theta_0(t)+\theta_1(t,\mathbf{x})$. Working in Newtonian gauge and focusing on superhorizon Fourier  modes we obtain the evolution equation
\begin{equation}
    \label{eq:Anharmonic13}
    \frac{d^2 \theta_1}{dy^2}+\frac{3}{2y}\frac{d\theta_1}{dy}+\theta_1(y)\cos{(\theta_0(y))}+2\Psi\sin{(\theta_0(y))}=0 \, ,
\end{equation}
whre $\Psi$ is the scalar curvature perturbation. This equation is solved numerically along with Eq.~\eqref{eq:Anharmonic1}.

The superhorizon isocurvature perturbation can be written as
\begin{equation}
    \label{eq:iso-pert}
    S_\phi(y,\zeta)=-\frac{3}{2y}\left(\frac{\delta\rho_\phi}{\Dot{\rho}_\phi}-\Psi y\right) \, ,
\end{equation}
where the background energy density $\rho_\phi$ is given in Eq.~(\ref{eq:Anharmonic2}), 
and the perturbed energy density $\delta\rho_\phi$ is
\begin{equation}
   \label{eq:Anharmonic14}
   \delta\rho_\phi(y,\zeta) = {m_{\phi}}^2 {f_{\phi}}^2 \left[\left(\frac{d\theta_0}{dy}\right) \left(\frac{d\theta_1}{dy}\right)-\Psi {\left(\frac{d\theta_0}{dy}\right)}^2+\sin{(\theta_0(y))}\theta_1(y)\right] \, .
\end{equation}
Using the numerical solutions $\theta_0(y)$ and $\theta_1(y)$, we obtain the evolution of isocurvature $S_\phi$, Eq.~\eqref{eq:iso-pert}. 
Since we are interested in the late time isocurvature, evaluated after cosmic expansion has redshifted $\theta_0(y)$ to small values, we use the harmonic approximation $\sin\theta_0(y)\simeq\theta_0(y)$ at $y_f\gg1$.

The anharmonic factor $\mathcal{F}_{\rm iso}(\zeta)$ can then be obtained according to 
\begin{equation}
    \label{eq:Anharmonic19}
    \mathcal{F}_{\textrm{iso}}(\zeta)\equiv \frac{S_{\phi}(y_f,\zeta)}{S_{\phi}(y_f,\zeta_{c})} \, ,
\end{equation}
with $\zeta_{c} \ll 1$ chosen to be $10^{-6}$.
The anharmonic factors for the energy density and isocurvature are shown in Fig.~\ref{fig:anharmonic}.

\bibliographystyle{jhep}
\bibliography{ref}
\end{document}